\documentclass[aps,prl,reprint,amsmath,amssymbols,superscriptaddress]{revtex4-1}

\usepackage{graphicx}
\usepackage{bm}
\usepackage{color}
\usepackage{amsmath}
\usepackage{amsfonts}
\usepackage{amssymb}

\newcommand\tcl[1]{\bgroup\color{red}\bfseries{#1}\egroup}
\definecolor{darkGreen}{rgb}{0,0.5,0}
\definecolor{darkBlue}{rgb}{0,0,0.5}
\definecolor{darkRed}{rgb}{0.5,0,0}

\newcommand{\sm}{G}
\newcommand{\zt}{\tilde{z}}  
\newcommand{\qv}{\mathbf{q}}
\newcommand{\Tr}{\text{Tr}}

\newcommand{\tp}{\tilde{p}}

\newcommand{\delete}[1]{{}}

\begin{document}
	\title{Jamming as a Multicritical Point}
	\author{Danilo B. Liarte}
	\email{dl778@cornell.edu}
	\affiliation{Laboratory of Atomic and Solid State Physics, Cornell University, Ithaca, NY, USA}
	\author{Xiaoming Mao}
	\affiliation{Department of Physics, University of Michigan, Ann Arbor, MI, USA}
	\author{Olaf Stenull}
	\affiliation{Department of Physics and Astronomy, University of Pennsylvania, Philadelphia, PA, USA}
	\author{T. C. Lubensky}
	\email{tom@physics.upenn.edu}
	\affiliation{Department of Physics and Astronomy, University of Pennsylvania, Philadelphia, PA, USA}
	
	\date{\today}
	
	\begin{abstract}
	The discontinuous jump in the bulk modulus $B$ at the jamming transition is a consequence of the formation of a critical contact network of spheres that resists compression. We introduce lattice models with underlying under-coordinated compression-resistant spring lattices to which next-nearest-neighbor springs can be added.  In these models, the jamming transition emerges as a kind of multicritical point terminating a line of rigidity-percolation transitions. Replacing the under-coordinated lattices with the critical network at jamming yields a faithful description of jamming and its relation to rigidity percolation.
	\end{abstract}
	
	\maketitle
	
	Jamming \cite{LiuNa2010a,LiuNa2010b} is now well-established as a phenomenon with a zero-temperature mechanical critical point that separates a state of free particles from one in which they collectively resist elastic distortions.  The jamming critical point ($J$) is, however, unusual in that it exhibits properties of both a first-order transition (with a discontinuous jump in the bulk modulus, $B$, and a second-order one (with a continuous growth of the shear moduli, $\sm$, from zero). This is in stark contrast to its cousin, the rigidity-percolation (RP) transition \cite{FengGar1985,Garboczi1985} in which both the bulk and shear moduli grow linearly from zero above the RP critical point (or line). The first-order jump in $B$ is a consequence of the formation of a critical network of contacts that resists compression.  This fact is the inspiration for our introduction of lattice models with sublattices that also resist compression. In our analysis of these models using effective medium theory (EMT) \cite{Elliott1974,FengGar1985} and numerical simulations, the jamming transition corresponds to a kind of multi-critical point at which a line (or surface) of RP transitions meets a line along which $B$ is nonzero.  
	\begin{figure}[!ht]
		\centering
		\includegraphics[width=\linewidth]{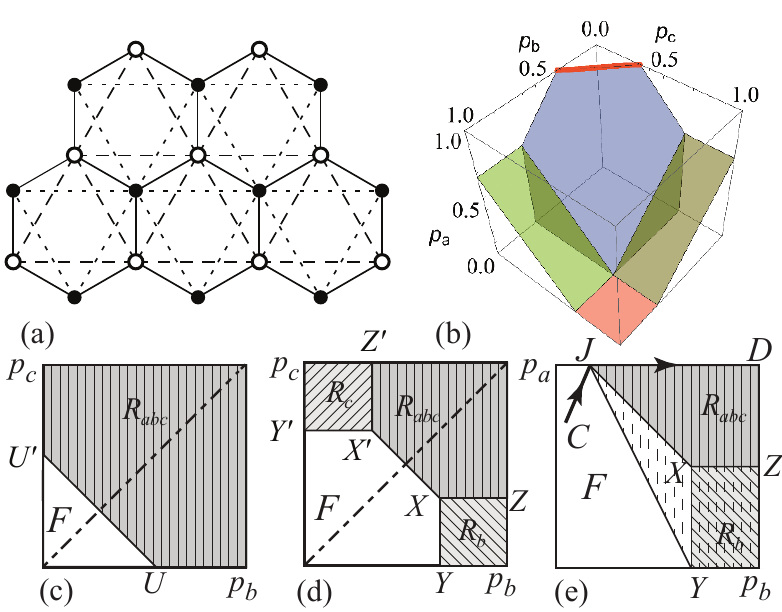}
		\caption{(a) 3-sublattice model showing NN (solid) and NNN (dashed and dotted) bonds, the latter of which connect sites in either of the triangular sublattices containing the first (black) or second (open) sites of the  honeycomb lattice. (b) 3D phase diagram showing the surfaces $S_{FRabc}$ (blue), $S_{Rb Rabc}$ (green), $S_{Rc Rabc}$ (khaki), $S_{FRb}$ (dark green), and $S_{F Rc}$ (dark khaki) and the jamming line $L_J$ (red); (c) to (e) 2D slices of the 3D diagram at (c) constant $2/3<p_a<1$, (d) constant $0<p_a<2/3$, and (e) at both $p_c=1/6$ at $p_b=p_c$, showing the $F$, $R_b$, $R_c$, and $R_{abc}$ phases.  $UU'$ is the jamming line when $p_a=1$ and an RP line when $p_a <1$.  $XX'$, $XY$, $X'Y'$, $JX$, and $JY$ are RP lines, and $XZ$ marks the transition form the $R_b$ to the $R_{abc}$ phase.  In (e) the region to the right of $JX$ is the $R_{abc}$ phase when $p_a = p_b$, and the line $JX$ is  not a part of the figure when $p_c=1/6$. CJD is a jamming path with a discontinuous jump in $B$}
		\label{fig:model}
	\end{figure}

	Our models begin with the under-coordinated honeycomb lattice in two dimensions (2D) or the diamond lattice in 3D, each consisting of sites connected by nearest-neighbor (NN) springs, with a non-vanishing bulk modulus but with vanishing shear moduli~\cite{LiarteLub2016}. Next-nearest-neighbor (NNN) springs are randomly added (as shown in Fig.~\ref{fig:model}(a)) leading to the phase diagrams shown in Fig.~\ref{fig:model}(b)-(e). At a critical concentration of NNN springs, the Maxwell rigidity criterion \cite{Maxwell1864}  is reached, the shear modulus begins to grow continuously from zero, and the bulk modulus begins to increase. This model mimics important aspects of jamming and the jamming transition, which is reached by increasing the volume fraction of spheres until they have a sufficient number of contacts to first resist compression, indicating a bulk modulus that is greater than zero.  The marginally jammed state that is formed is an analog of the honeycomb or diamond lattice in our model.  Further compression of the jammed lattice increases the number of contacts and produces an increase in the shear moduli from zero.  This is the analog of adding NNN bonds in our models. Our model differs from jamming in that sites in the former are fixed on a periodic lattice whereas those in the latter are off lattice and change positions with compression. In addition, the bulk modulus in our models remains nonzero below the jamming transition as long as the NN bonds are occupied with unit probability.  Our approach, however, can be applied to any lattice that has an under- or critically coordinated sublattice with  nonzero $B$, such as that discussed at the end of this paper and in the SI.
		
	Our model exploits the fact that both the honeycomb and the diamond lattices with NNN bonds can be divided into three independent bond lattices, each sharing the sites of the original NN lattice: the original NN lattice (the $a$-lattice), and two independent NNN lattices (the $b$ and $c$ lattices) with sites, respectively, on one or the other site-sublattices of the $a$-lattice (see Fig.~\ref{fig:model}(a)). Population of the bonds of these lattices with springs of spring constant $k$ with probabilities $p_a$, $p_b$, and $p_c$ gives rise to EMT spring constants $k_a$, $k_b$, and $k_c$, respectively. In what follows, we will focus on the $2D$ case, though most of the results we present apply to the diamond lattice as well.
	
	The full $3D$ EMT phase diagram, depicted in Fig.~\ref{fig:model}(b), in the space defined by $(p_a,p_b,p_c)$ shows four distinct phases: a floppy phase $F$, in which $B$ and $\sm$ are zero at zero frequency, and three rigid phases with $B>0$ and $\sm>0$: $R_b$, in which only $k_b>0$; $R_c$ in which only $k_c>0$; and $R_{abc}$, in which $k_a$, $ k_b$, and $k_c$ are all nonzero. It addition, it shows boundary surfaces $S_{AB}$ where phases $A$ and $B$ meet, lines $L_{ABC}$ where phases $A$, $B$, and $C$ meet, and the jamming line $L_J$ (in red) where $S_{FRabc}$ meets the plane $p_a=1$. Figures~\ref{fig:model}(c), (d), and (e) depict various 2D slices.  In (c) the line $U^\prime U$ is $L_J$ when $p_a=1$. In (e), $J$ is the point on $L_J$ when $p_c = 1/6$. Surfaces $S_{FR}$ for $R$ equal to $R_b$, $R_c$, or $R_{abc}$, and lines $XY$, $XX^\prime$, ${Y^\prime}X^\prime$, $JX$, and $JY$ correspond to $RP$ transitions; and surfaces $S_{RRabc}$, with $R=R_b$ or $R_c$, and line $XZ$ represent transitions in which $k_a$ develops a non-zero value when $k_b$ or $k_c$ is already non-zero.  In (e), $J$, viewed from the $F$ phase, is a critical endpoint \cite{ChaikinLub1995} where the second-order RP line $JY$ meets the first-order line at $p_a =1$. In what follows, we will focus on the vicinity of the points $J$ and $X$ in the 2D slices.
	
	As shown in previous studies (see e.g.~\cite{FengGar1985,SchwartzSen1985,JacobsThor1995}), the EMT provides accurate but not exact estimates of elastic moduli and phase boundaries. In particular, it does not incorporate redundant bonds \cite{JacobsThor1995} that lead to over- and under-constrained regions in randomly-diluted samples. Our results agree with this previous work (see Fig.~\ref{fig:plots}): simulations and EMT track each other closely, but with larger deviations near rigidity transitions and particularly near point $X$ where simulations do not show discontinuous slope changes predicted by the EMT (See SI).
	
		\begin{figure}[!ht]
		\includegraphics[width=\linewidth]{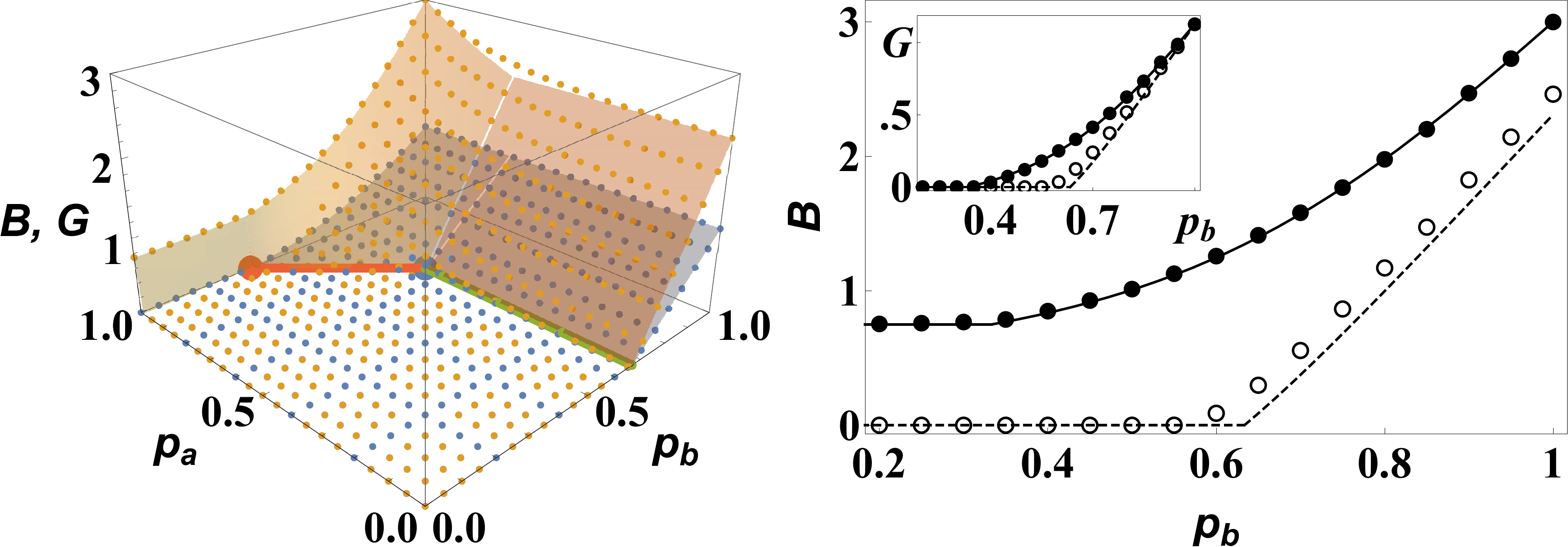}
		\caption{Left: Simulations (points) and EMT solutions (surfaces) for $B$ (yellow) and $G$ (blue) as a function of $p_a$ and $p_b$ for $p_c=0$. Red and green lines correspond to $JX$ and $XY$ on Fig~\ref{fig:model}e, respectively. Right: $B$ and $G$ (inset) as a function of $p_b$ from EMT (lines) and simulations (circles) for $p_c = 0$ and $p_a=1$ (filled circles) and $p_a=0.7$ (open circles).
			\label{fig:plots}}
		\end{figure}

	In Fig.~\ref{fig:model}(e), $k_a$, and thus $B$, is nonzero along the line $p_a=1$, but $k_b$ and $k_c$ are (for both $p_b=p_c$ and $p_c = 1/6$) zero along this line for $p_b$ less than or equal to its value $p_b^J$ at $J$.  Thus, $G$ but not $B$ approaches zero as $J$ is approached along not only the line $p_a=1$, but along any line approaching $J$ from from the rigid side.  On the other hand if $J$ is approached from the floppy side along any path (e.g., CJD in Fig.~\ref{fig:model}(e)) other than $p_a = 1$, $B$ will undergo a discontinuous change at $L_J$ as in jamming. We argue that a path with $p_a <1$ until $J$ is reached followed by a path along $p_a =1$ for $p_b >p_b^J$ faithfully represents the jamming transition. If springs are removed randomly from a jammed lattice at $J$, it immediately loses its rigidity.  This also takes place in our model if we allow removal of springs from the $a$-lattice as well as the $b$ and $c$ lattices, i.e., follow a path in $F$ in which $p_a<1$ until $L_J$ is reached. The jamming line at $p_a=1$ terminates an RP surface ($S_{FRabc}$) across which all effective spring constants, and thus both $B$ and $\sm$, grow linearly with distance from it.
	
	EMTs also yield information about finite frequency behavior \cite{Elliott1974,WyartMah2008,Wyart2010,MaoLub2011a,BroederszMac2011,MaoLub2013a,DuringWya2013} with the inclusion of inertia of mass points and/or viscous friction with a background fluid \cite{YuchtBroe2013}. In our case, the former yields densities of states that scale like those near jamming, and the latter lead to renormalized shear and bulk viscosities in the floppy regime, the former of which diverge as $|\Delta \tilde{p}|^{-1}$ at the $S_{RP}$'s and along $L_J$, and the latter of which also diverge as $|\Delta \tilde{p}|^{-1}$ at the $S_{RP}$'s but as $|\Delta \tilde{p}|^{-2}$ along paths terminating at $L_J$.
			
	Our EMT replaces randomly placed springs with spring constant $k=1$ in the three lattices with homogeneously placed ones with respective effective spring constants $k_a$, $k_b$, and $k_c$ such that the average scattering from any given spring in the effective background medium is zero.  The EMT equations are then
	\begin{eqnarray}
		k_\alpha (\omega) & = &[p_\alpha -h_\alpha (\omega)]/[1-h_\alpha (\omega)] ,\qquad \alpha = a,b,c,
		\label{eq:EMT2}\\
		h_{\alpha} (\omega) &= &\frac{1}{\zt_\alpha N_c}
		\sum_{\qv} \Tr\,k_\alpha (\omega) K_\alpha (\qv)\mathcal{G}(\qv,\omega) ,                                    
		\label{eq:EMT1}
	\end{eqnarray}
	where $\mathcal{G}(\qv,\omega)=[\sum_{\beta}k_\beta(\omega) K_\beta(\qv)-w(\omega ) I]^{-1}$ the lattice Green's function, $N_c$ the number of unit cells, $\zt_{\alpha}$ (= 3 for all $\alpha$ in the honeycomb lattice) the number of bonds per unit cell in lattice $\alpha$ ($= a,b,c$), $K_\alpha (\qv)$ is the $\alpha$-lattice normalized stiffness matrix, and $w(\omega) = \omega^2 + i \gamma \omega$, where $\omega$ is the frequency, $\gamma$ is the drag coefficient, and the mass is set to one. As discussed in the SI, evaluation of $h_\alpha$ in the limit $k_b$, $k_c$, and $w$ tend to zero requires some care because $K_a$ has a zero eigenvalue at every $\qv$. The $k_\alpha (\omega)$ are determined by the self-consistent solution to Eqs.~(\ref{eq:EMT2}) and (\ref{eq:EMT1}). In the zero-frequency limit ($w(\omega)\rightarrow 0$), $k_\alpha \equiv k_\alpha (\omega=0)=0$ when $p_\alpha = h_\alpha (\omega=0)\equiv h_\alpha$, $k_\alpha = 1$ when $p_\alpha = 1$, and $0\leq k_\alpha \leq 1$ for $h_\alpha \leq p_\alpha \leq 1$.  As we shall see, $k_\alpha$ vanishes as $w(\omega) \rightarrow 0$ when $p_\alpha < h_\alpha$
	
	It follows from Eq.~(\ref{eq:EMT1}) that the $h_{\alpha}$'s satisfy the sum rule
	\begin{equation}
	\sum_{\alpha} \zt_\alpha h_{\alpha}(\omega) 
		= m D  [1 + (w(\omega)/N_c)\sum_{\qv}\Tr \mathcal{G}(\qv,\omega)] ,
	\label{eq:sumrule}
	\end{equation}
	where $D$ is the spatial dimension and $m=2$ is the number of sites per unit cell in the honeycomb and diamond lattices.  Equation (\ref{eq:sumrule}) along with the results of Eq.~(\ref{eq:EMT2}) that $h_\alpha =p_\alpha$ when $k_\alpha = 0$ yield the Maxwell condition for marginal stability on the $S_{FRabc}$ surface or on the jamming line at $\omega=0$:
	\begin{equation}
	\zt_a p_a  +\zt_b(p_b + p_c) = m D.
	\label{eq:Maxwell} 
	\end{equation}
	The surfaces $S_{FRb}$ and $S_{FRc}$ signal the onset of rigidity of the $b$ and $c$ lattices individually, in which case, $k_a$ and $k_b$ ($k_c$)  adopt the vanishing solutions  to Eq.~(\ref{eq:EMT1}). In this case, the rigid $b$ ($c$) lattice is triangular and has only one site per unit cell, and $h_b =D/\zt_b=2/3$ throughout the $R_b$ phase, and  similarly for $h_c$.  At $S_{RbRabc}$, $k_a$ and $k_c$ first adopt non-zero solutions to Eqs.~(\ref{eq:EMT1}) and (\ref{eq:EMT2}), and $h_a = p_a$ and $h_c=p_c$ to yield  $\zt_a p_a + \zt_b p_c = (m-1)D$ on $S_{RbRabc}$.
	
	We will now focus on critical points and lines in Figs.~\ref{fig:model}(d) and~\ref{fig:model}(e). As noted above, $J$ marks the jamming  point and $X$ the critical point where $F$, $R_{abc}$, and $R_b$ meet. At fixed $p_c$, $J = (1,p_b^J,p_c)$, where $p_b^J= (1/3)-p_c$, and $X=(2/3-p_c,2/3,p_c)$ (for $0<p_c<2/3$).  At $p_b=p_c$, $J=(1,1/6,1/6)$ and $X=(1/2,2/3,1/6)$.  Figure \ref{fig:model}(e) shows phase-diagram slices for $p_c=1/6$ and for $p_b = p_c$. The lines $JX$ and $JY$ satisfy the equation
	\begin{equation}
	\Delta \tp \equiv \Delta p_b^J-\nu \Delta p_a^J = 0,
	\label{eq:Dtp}
	\end{equation}
	where $\Delta p_b^J=p_b-p_b^J$, $\Delta p_a^J =(1-p_a)>0$, and the inverse slope, is $\nu=\nu_X=1$ for the line $JX$ at fixed $p_c=1/6$ and $\nu=\nu_Y=1/2$ for the line $JY$ and $p_c= p_b$.
	
	Along the $F$-$R_{abc}$ lines $JX$ or $JY$, all effective spring constants (on bonds with non-zero occupation probability), and thus all elastic moduli, grow linearly with $\Delta \tp$, and along the $F$-$R_b$ line, $k_b$ grows linearly with $\Delta p_b = p_b -2/3$:
	\begin{equation}
	k_r^{JV}=c_r^{JV} [\Delta \tp], \qquad  k_b^{XY} = c_b^{XY} [\Delta p_b] ,
	\label{eq:kRp}
	\end{equation}
	where $[\phi]=(\phi+|\phi|)/2$, $r=a,b$ and $V=X,Y$, and $c_r^{JV}$ varies with position along $JV$.  Along the line $p_a=1$, $k_a$ is exactly equal to one.  Near $J$, $k_b$ maintains its form of Eq.~(\ref{eq:kRp}), but $k_a$ has to vanish on $JV$ and equal one at $p_a=1$.  This is accomplished within the EMT by
	\begin{equation}
	k_b^J=\frac{[\Delta \tp]}{s + \nu c_J}; \,
	k_a^J = \frac{c_J k_B^J}{c_Jk_B^J+  \Delta p_a} \rightarrow
	\frac{c_J\Delta\tp}{c_J \Delta p_b + s \Delta p_a} ,
	\label{eq:kaJ}
	\end{equation}
	where $s=1-p_b^J$.  When $\Delta p_a = 0$, $k_a^J = 1$ as required.  Also $k_a^J$ clearly vanishes along $JV$ where $\Delta \tp = 0$. The elastic moduli of the honeycomb lattice in terms of the $k$'s are $G = r_b k_b +r_c k_c$ and $B= s_a k_a + s_b k_b + s_ck_c$, where $r_b=r_c=9/8$, $s_a=3/4$, $s_b=s_c=9/4$, and as advertised, $G$ vanishes linearly with $\Delta\tp$. The value of $k_a$ and thus of $B$ depends on the path to the jamming point as can be seen by putting $\Delta p_b = \nu^{\prime} \Delta p_a$ in Eq.~(\ref{eq:kaJ}) with $\nu^{\prime} >\nu$: $k_a^J= c_J(\nu^{\prime}- \nu)/(c_J\nu^{\prime}+s)$.  The ratio $G/B$ approaches zero and the Poisson ratio $\sigma$ approaches its limit value of one along all paths to $J$. $G/B$ reaches a value along the $RP$ line $JY$ increasing from zero at $J$ to a maximum of $1/2$ at $Y$.  These results are similar to those in Ref.~\cite{GoodrichNag2014,GoodrichNag2015}.
	
	We now turn to behavior in the vicinity of $X$. The EMT solution at $w=0$ is
	\begin{equation}
	k_b^X =[\Delta \tp_{ab}^X]/s_b \qquad \text{and}\qquad
	k_a^X = k_b^X [\Delta p_a^X]/c_X ,
	\label{eq:kX}
	\end{equation}
	where $\Delta \tilde{p}_{ab}^X = \Delta p_b^X + \nu_X [\Delta p_a^X]$, $\Delta p_b^X = p_b - p_b^X$, $\Delta p_a^X = p_a- p_a^X$, $c_X \approx 0.1$ (evaluated numerically), and $s_b = 1-p_b^X$. These equations encode all of the phase boundaries incident at $X$:  $\Delta \tilde{p}_{ab}^X$ is equal to $\Delta p_b^X$ when $\Delta p_a^X<0$ and to $\Delta \tp^X = \Delta p_b^X + \nu_X \Delta p_a^X$ when $\Delta p_a^X>0$ so that $k_b^X = 0$ for $\Delta p_a^X<0$ and $\Delta p_b^X <0$ and for $\Delta \tp^X <0$ and $\Delta p_a^X >0$.  The result is that $k_b^X>0$ in the $R_b$ and $R_{abc}$ phases in Fig.~\ref{fig:model} and that $k_a^X$ is nonzero only in the $R_{abc}$ phases of that figure. We have calculated the bulk and shear moduli by numerical solution of the EMT equations for the $k_\alpha$'s and by their direct evaluation on our random lattices. The two solutions are nearly identical over most of phase space as seen in Fig.~\ref{fig:plots}.  The simulations, however, do not show the sharp changes near $X$ that the EMT does.

\begin{figure}[!ht]
	\centering
	\includegraphics[width=\linewidth]{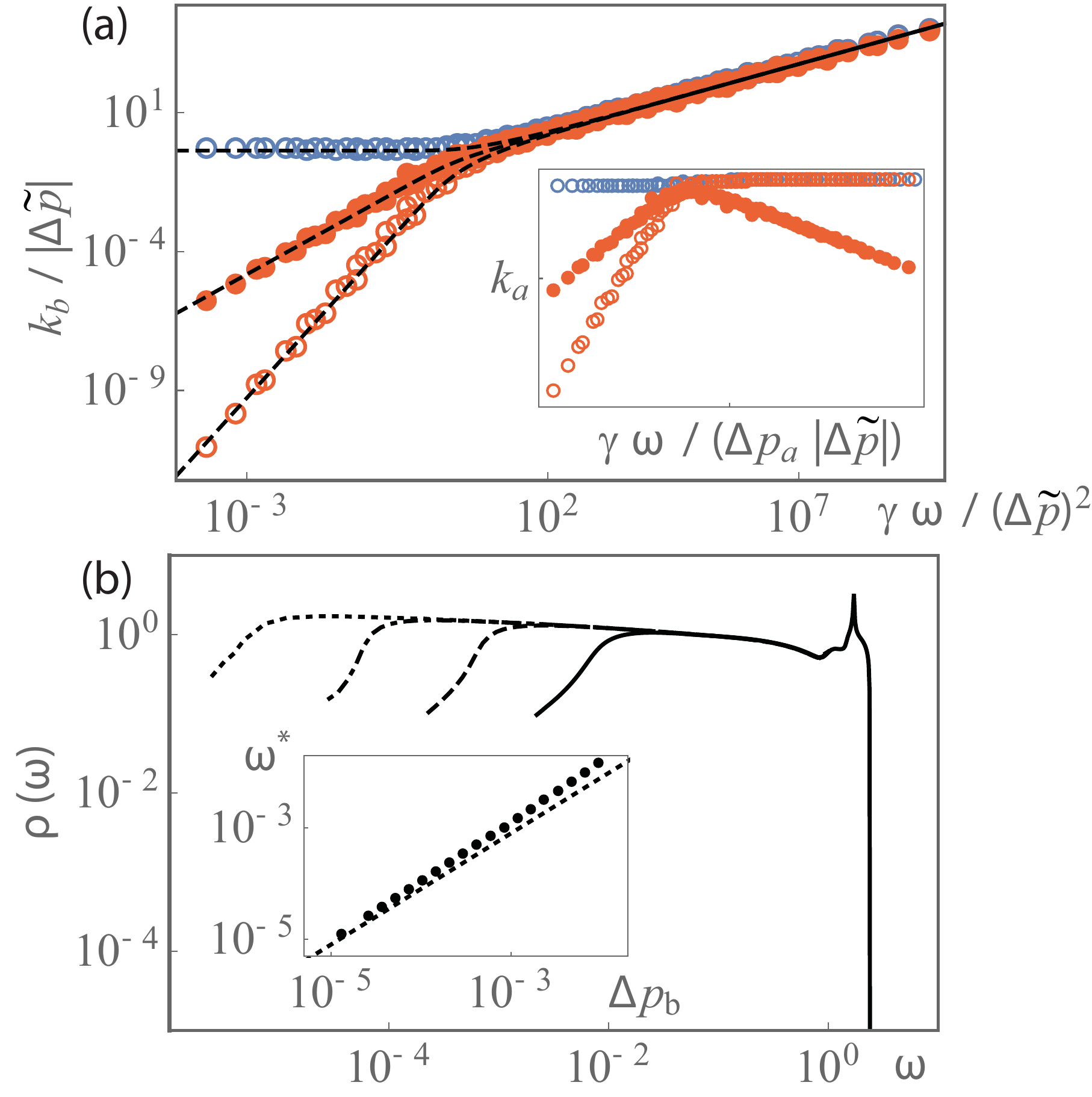}
	\caption{(a) $k_b / |\Delta \tilde{p}|$ as a function of $\gamma \omega / |\Delta \tilde{p}|^2$ in the low-frequency limit $w= i \gamma \omega$.  Blue (red) circles: numerical solutions to full EMT equations for approach to jamming in the rigid (floppy) phase; black dashed line: Asymptotic  solutions [Eq.~\eqref{eq:kb-omega}] near jamming critical point; Hollow circles: $\text{Re} k_b/|\Delta \tilde{p}|$; Filled circles: $-\text{Im} k_b/|\Delta \tilde{p}|$, which is independent of the  sign of $|\Delta \tilde{p}|$. Inset: $k_a$ as a function of $\gamma \omega/(\Delta p_a|\Delta \tp | )$. (b) Density of States $\rho(\omega)$ for $p_a = 1$ and $\Delta \tilde{p}=\Delta p_b = 10^{-2}$ (solid lines), $10^{-3}$ (dashed), $10^{-4}$ (dot-dashed) and $10^{-5}$ (dotted); Inset: Linear behavior of crossover frequency $\omega^* (\Delta p_b)$.}
	\label{fig:dynamics}
	\end{figure}

	Equation (\ref{eq:EMT1}) provides dynamical as well as static information, allowing us to calculate the frequency-dependent effective spring constants in the floppy region.  Of particular interest is the approach to the jamming point. In the case of $p_b=p_c$, the results (in agreement with Ref.~\cite{YuchtBroe2013} for $k_b$) are
	\begin{align}
	k_b & = \frac{1}{2(s+\nu c)}\left[\Delta \tp + \sqrt{|\Delta \tp|^2 -4(s+ \nu c_J) v_b w(\omega)}\right],	\label{eq:kb-omega} \\
	& \approx 
	\frac{[\Delta \tp]}{s+\nu c_J} -\frac{v_b w}{|\Delta \tp|} ,\qquad \text{when}\,\,\,\frac{v_b w}{|\Delta \tp|^2} \ll 1,
	\end{align}
	and
	\begin{align}
	k_a & =\frac{k_b}{k_b +( \Delta p_a/c_J)} \\
	& \xrightarrow{\Delta \tp<0} 
	\frac{v_b w}{v_b w + (\Delta p_a |\Delta \tp|/c)} 
	\approx \frac{c_J v_b w}{\Delta p_a |\Delta \tp|},\label{eq:ka-omega}
	\end{align}
	Thus on paths approaching $J$ in the low-frequency limit when $w= i \gamma \omega$, $k_b$ diverges as $|i \gamma \omega \Delta \tp|^{-1}$, but $k_a$ diverges as $i \gamma \omega |\Delta p_a \Delta \tp|^{-1}$, implying that the shear viscosity diverges as  $|\Delta \tp|^{-1}$, but the bulk modulus viscosity diverges as  $|\Delta \tp|^{-2}$ . The scaling of $k_b$ [Fig.~\ref{fig:dynamics}(a)] is consistent with results for the shear modulus of soft sphere packings near jamming~\cite{tighe11}. When $\gamma =0$ and  $w=\omega^2$, our calculations yield a density of states that is nearly constant at small $\omega$ [Fig.~\ref{fig:dynamics}(b)], down to a crossover frequency $\omega^*$ that scales as $\Delta \tilde{p}$ (see inset), as in jamming~\footnote{Our calculations show that the linear crossover from elasticity to isostaticity is akin to jamming \emph{and} rigidity percolation. As shown in the SI, jamming and rigidity percolation can display different Debye behavior for the density of states at very low frequencies.}.

	As noted earlier, in our model, $k_a$, and thus $B$, is nonzero in the floppy region when $p_a = 1$. In the jamming protocol, $B$ is zero in the floppy phase and jumps discontinuously at $J$ with the formation of a random marginally stable lattice with a single state of self stress \cite{Calladine1978,GoodrichNag2012} that resists increase in pressure of volume fraction.  As volume fraction is increased, more links form, inviting us to model jamming starting with the lattice at $J$, which is now critically rather than under coordinated with $\tilde{z}_a=D$ ($\tilde{z}_a$ is half the coordination number), as the analog of the $a$ lattice and identifying ``unoccupied bonds'' between pairs of close but not touching spheres as the $b$ lattice.  Ideally this $b$ lattice would contain a sufficient number of bonds that it would by itself be mechanically stable if all of these bonds were occupied with springs. We can now use the random-lattice EMT of Refs. \cite{WyartMah2008,Wyart2010,DuringWya2013}, modified to treat lattices $a$ and $b$ separately.  The result is a phase diagram [See SI] in the $p_a$-$p_b$ space identical to that of Fig.~\ref{fig:model}(e) but with the point $J$ moved to the upper left hand corner: $J=(1,0)$ and the point $Y$ moved to $Y=(0,D/\tilde{z}_b)$.  The path to jamming, which involves first the creation of lattice $a$, is thus along the line $p_b=0$ until $J$ is reached. As more springs are added, the path follows the line $p_a=1$.  Of course, different paths can be followed, most of which will intersect the $RP$ line $J$-$Y$ \cite{FengGar1985,GoodrichNag2015}.  For example, all paths starting from a point in the jammed phase along $p_a=1$ in which springs are randomly removed from both $a$ and $b$ sublattices cross the $RP$ line. The EMT equations are identical in form to Eqs.~(\ref{eq:EMT1}) and (\ref{eq:EMT2}), but with only two sublattices and Eq.~(\ref{eq:Maxwell}) replaced by $\tilde{z}_a p_a + \tilde{z}_b p_b = D$, where $\tilde{z}_a=D$.  Near $J$, $k_a$ and $k_b$ obey Eqs.~(\ref{eq:kRp}), (\ref{eq:kaJ}), and (\ref{eq:ka-omega}) to (\ref{eq:kb-omega}) with $p_b^J = 0$ and $s=1$. See the Supplementary Information for more detail.
	
	Our model features a second-order RP line meeting a first-order $B>0$ line.  Possible procedures for producing similar features in jammed systems include targeted selective pruning \cite{GoodrichNag2015,EllenbroekHec2015} or dividing bonds into those present in the marginal network at jamming and those added later followed by removal of the former and latter with respective probabilities $p_a$ ad $p_b$.
	
	In this article, we introduced and analyzed, using effective medium theory and numerical simulations, a lattice model for jamming that captures the essential features of the jamming transition, which emerges as a critical endpoint in which a second-order rigidity percolation line meets a line in which there is a discontinuous jump in the bulk modulus from a non-rigid phase.
	
\begin{acknowledgments}	
	{This work was supported in part by NSF MRSEC/DMR-1720530 (TCL and OS), NSF DMR-1719490 (DBL), and 	
		NSF DMR-1609051 (XM).}
\end{acknowledgments}


%

\newpage
\clearpage
\begin{widetext}
\begin{center}
\textbf{\large Supplemental Material to: \\ Jamming as a multicritical point}
\end{center}

{\small
	In this text, we provide additional details of some of the main analytical and numerical results
	of the paper.
	First, we introduce the honeycomb and diamond lattice models, and show some of the phonon
	and elastic properties of the model without disorder.
	Second, we introduce the Coherent Potential Approximation and present a brief derivation of
	the effective-medium self-consistent equations.
	We then give details of calculations of the static and dynamic solutions near a few particular 
	regions of interest, such as the jamming line, in the phase diagram.
	Third, we make a few remarks about our numerical simulations, and present simulation and
	effective-medium theory results for the elastic moduli of the 3D diamond lattice.
	Fourth, we briefly discuss a generalization of our model in which $B$ is nonzero only in the
	jamming phase.}
\vspace{0.5cm}
\end{widetext}

\setcounter{equation}{0}
\setcounter{figure}{0}
\setcounter{table}{0}
\setcounter{page}{1}
\makeatletter
\renewcommand{\theequation}{S\arabic{equation}}
\renewcommand{\thefigure}{S\arabic{figure}}
\renewcommand{\bibnumfmt}[1]{[S#1]}
\renewcommand{\citenumfont}[1]{S#1}

\section{Phonon and elastic properties}
	\label{sec:phonons-elastic}
	
	Here we give additional details of the phonon and elastic properties of the honeycomb
	and diamond-lattice models.
	In what follows, we briefly review the phase diagram of our model, write down equations
	for the interaction energy, dynamical matrix and elastic moduli, and discuss dispersion
	relations.

	\begin{figure}[!ht]
		\begin{minipage}[t]{0.48\linewidth}
			\centering (a) \par\smallskip
			\includegraphics[width=0.8\linewidth]{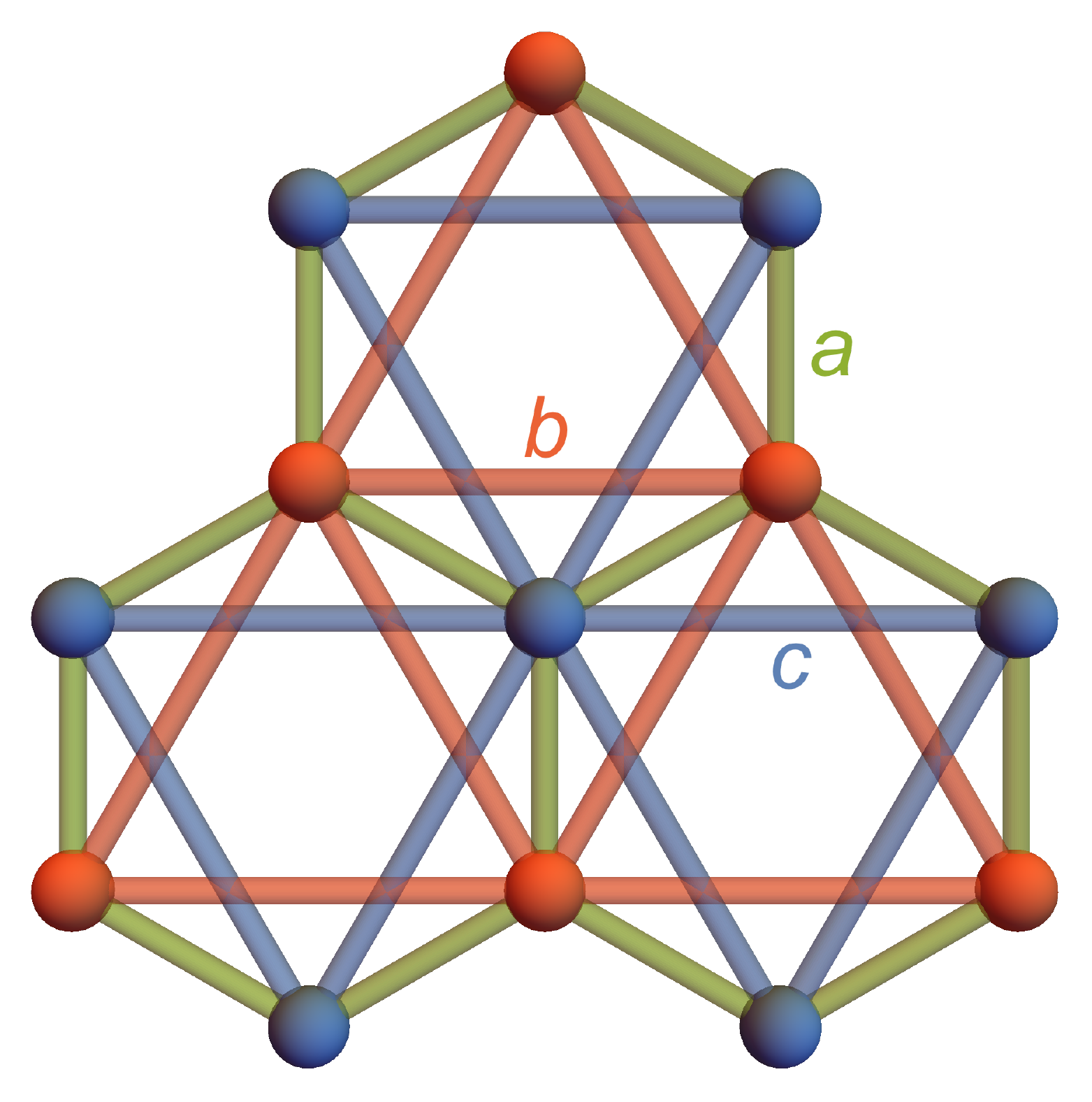}
		\end{minipage}
		\hspace{0.1cm}
		\begin{minipage}[t]{0.48\linewidth}
			\centering (b) \par\smallskip
			\includegraphics[width=0.8\linewidth]{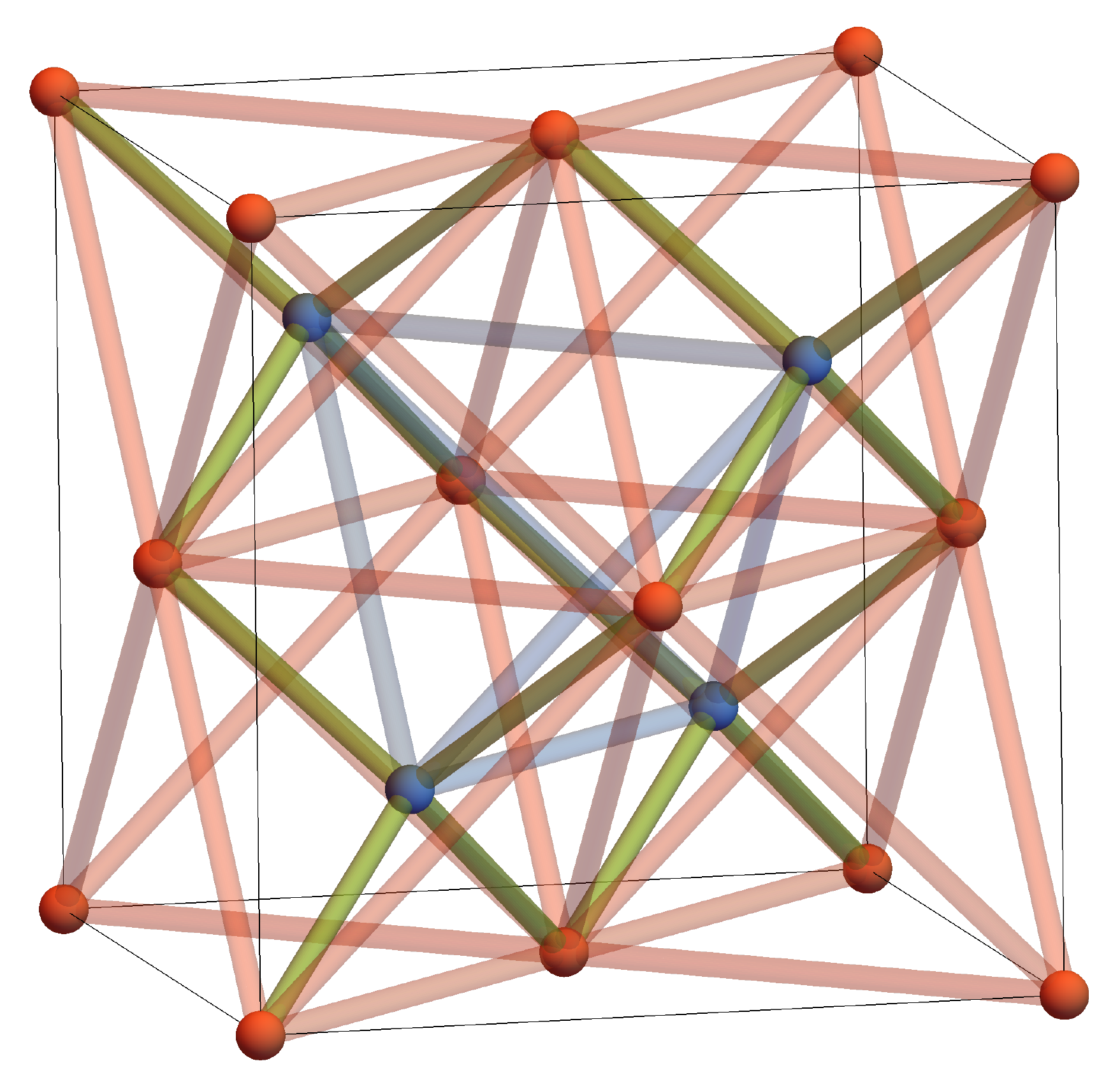}
		\end{minipage}
		\hspace{0.1cm}
		\begin{minipage}[t]{\linewidth}
			\centering (c) \par\smallskip
			\includegraphics[width=0.5\linewidth]{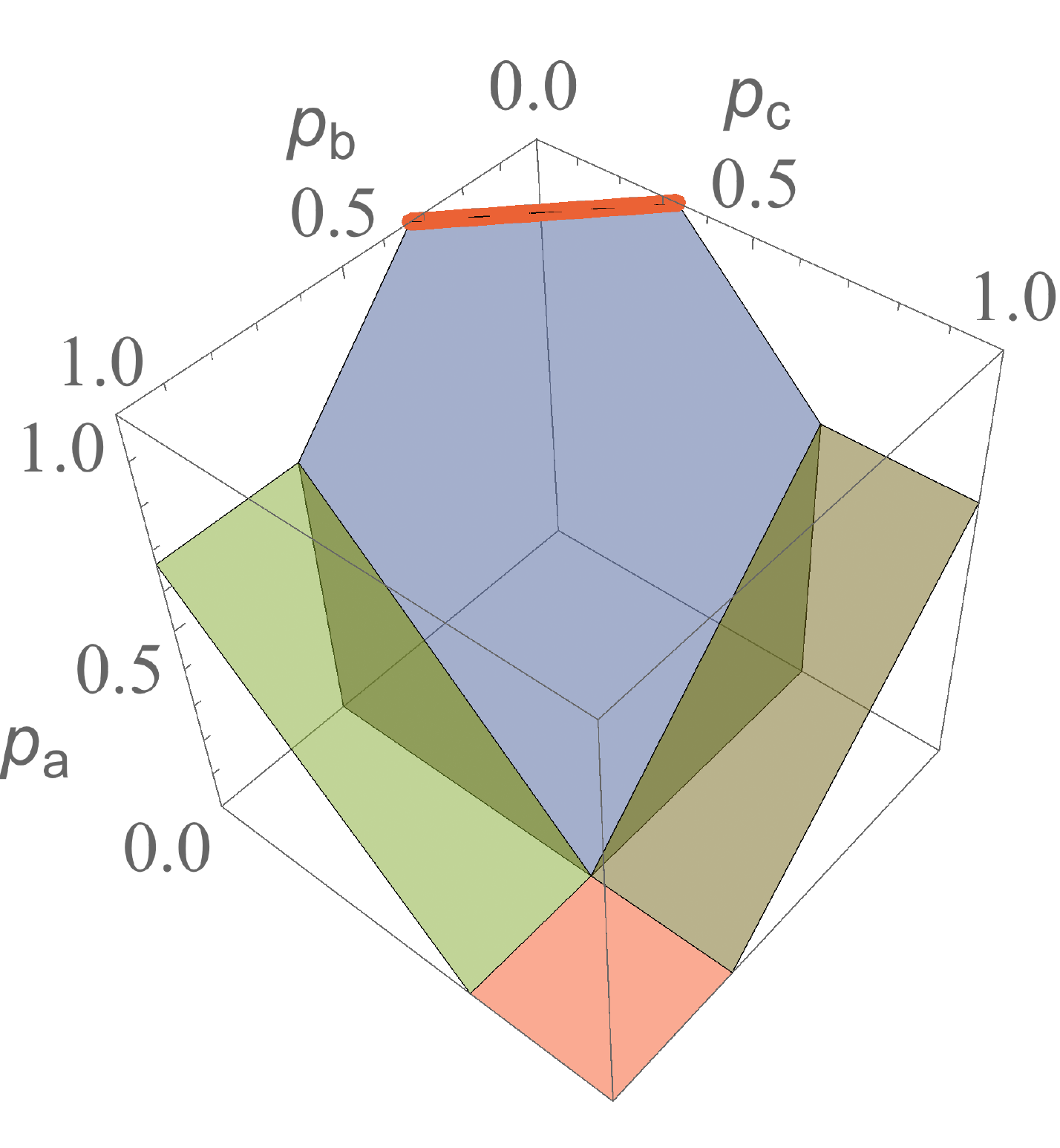}
		\end{minipage}
		\caption{
			Illustration of the lattice structure of the honeycomb (a) and diamond (b)
			lattice models.
			Green, red and blue lines represent bonds in the $a$, $b$ and $c$ sub-lattices,
			and are populated with probabilities $p_a$, $p_b$ and $p_c$, respectively.
			(c) Global phase diagram of the HLM in $p_a \times p_b \times p_c$ space.
			The red line at $p_a=1$ is the jamming multi-critical line.
			\label{fig:HDLM}
		}
	\end{figure}
				
	Figure~\ref{fig:HDLM} describes the three-sub-lattice model considered in this paper.
	(a) and (b) show the 2D honeycomb and 3D diamond lattice models (HLM and DLM,
	respectively), where green, red and blue lines represent bonds in sub-lattice  $a$, $b$
	and $c$, respectively.
	We model a spring constant $k^\prime$ of sub-lattice $\alpha \in \{a,b,c\}$ as an
	independent random variable with probability distribution
	$P(k^\prime) = p_\alpha \delta( k^\prime -1 ) + (1-p_\alpha) \delta( k^\prime )$.
	Note that the HLM has a minor issue associated with the fact that some of the bonds
	connecting next-nearest-neighbor pairs of sites can intersect, depending on the values
	of $p_a$, $p_b$ and $p_c$.
	This bond-crossing property, which becomes a nuisance if one wants to study
	2D-printed versions of this system, can be avoided by setting $p_b$ or $p_c$ to zero.
	On the other hand, the 3D DLM does not have this property.
	In (c), we show the global effective-medium theory (EMT) phase diagram of the HLM
	in $p_a \times p_b \times p_c$ space.
	The DLM phase diagram is qualitatively identical.
	All sub-lattices are rigid in the region above the blue pentagon and the two green and
	the red rectangle.
	Only the $b$ ($c$) sub-lattice is rigid in the region below the green rectangles, i.e. the
	effective energy constant $k_b (k_c) > 0$, whereas $k_a=k_c (k_b)=0$.
	In the red rectangle region, the $b$ and $c$ sub-lattices are rigid and
	disconnected ($k_b, k_c > 0$ whereas $k_a=0$).
	All sub-lattices are ``floppy'' in the remaining region of the phase diagram.
	The red line at $p_a=1$ is the jamming line.
	
	We consider harmonic elastic interactions between nearest (NN) and next-nearest
	(NNN) neighbor pairs of sites of the honeycomb and diamond lattices (see
	Fig~\ref{fig:HDLM}).
	The interaction energy can be written as
	\begin{align}
		E
			= \sum_{\alpha\in\{a, b, c\}} \frac{k_\alpha}{2} \sum_{\{i,j\} \, \in \, C_\alpha }
				g_{ij}^\alpha \left[ \left( \bm{u}_j - \bm{u}_i\right) \cdot
				\bm{\hat{r}}_{ij} \right]^2,
		\label{eq:energyModel}
	\end{align}
	where $\bm{u}_i$ is a displacement vector,
	$\bm{\hat{r}}_{ij} = ( \bm{r}_j - \bm{r}_i ) / |\bm{r}_j - \bm{r}_i|$, with $\bm{r}_i$ giving
	the position of site $i$ in reference space, and $C_\alpha$ is a set of neighbor pairs
	of sites for sub-lattice $\alpha$.
	In the simulations (see Section~\ref{sec:simulation}), $k_a=k_b=k_c=1$, and
	$g_{ij}^\alpha=1$ and $0$ with probabilities $p_\alpha$ and $1-p_\alpha$,
	respectively; i.e. bonds are occupied by springs with probabilities $p_a$, $p_b$ and
	$p_c$.
	In the effective-medium theory (see Section~\ref{sec:cpa}), $g_{ij}^\alpha=1 \,
	\forall i$, $j \in C_\alpha$, and all bonds are occupied by effective springs with
	$k_\alpha$ determined by a set of self-consistent equations that depend on $p_a$,
	$p_b$ and $p_c$.
	
	Equation~\eqref{eq:energyModel} can be written as a quadratic form in Fourier
	space:
	\begin{equation}
		E
			= \frac{1}{2N_c^2} \sum_{\bm{q}, \bm{q}^\prime} \bm{u}({\bm{q}}) \cdot
				D({-\bm{q}, \bm{q}^\prime}) \cdot \bm{u}({\bm{q}^\prime}),
	\end{equation}
	where $N_c$ is the number of cells, $\bm{u}({\bm{q}})$ is the Fourier transform
	of $\bm{u}(\bm{r})$, and the translationally-invariant dynamical matrix can be
	written as
	\begin{equation}
		D({-\bm{q}, \bm{q}^\prime})
			= N_c \delta_{\bm{q}, \bm{q}^\prime} D({\bm{q}}),
	\end{equation}
	with
	\begin{equation}
		D({\bm{q}})
			= \sum_{\alpha\in\{a,b,c\}} k_\alpha K_{\alpha}(\bm{q}),
	\end{equation}
	where $K_{\alpha}({\bm{q}})$ is the $\alpha$-lattice stiffness matrix,
	\begin{equation}
		K_{\alpha}({\bm{q}})
			= \sum_{n=1}^{\tilde{z}_\alpha} \bm{B}_{n}^{\alpha}({\bm{q}})
				\bm{B}_{n}^{\alpha}({-\bm{q}}),
	\end{equation}
	$\tilde{z}_\alpha$ is the number of bonds per unit cell of sub-lattice $\alpha$, and
	\begin{align}
		& \bm{B}_{n}^a(\bm{q})
			= \{ -\bm{e}_n, e^{-i\bm{q} \cdot \bm{f}_n} \bm{e}_n\},
				\quad \text{for } 1 \leq n \leq \tilde{z}_a, \label{eq:Ba} \\
		& \bm{B}_{n}^b(\bm{q})
			= \{ \left(e^{-i \bm{q} \cdot \bm{a}_n} - 1 \right) \hat{\bm{a}}_n, \bm{0}\},
				\quad \text{for } 1 \leq n \leq \tilde{z}_b, \label{eq:Bb} \\
		& \bm{B}_{n}^c(\bm{q})
			= \{ \bm{0}, \left(e^{-i \bm{q} \cdot \bm{a}_n} - 1 \right) \hat{\bm{a}}_n \},
				\quad \text{for } 1 \leq n \leq \tilde{z}_c, \label{eq:Bc}
	\end{align}
	with $\bm{0}$ denoting a $D$-dimensional null vector, and
	$\hat{\bm{a}}=\bm{a}/|\bm{a}|$.
	Table~\ref{tab:lattice} lists particular forms for the set of vectors used in our
	calculations.
	$\bm{a}$ are lattice vectors connecting pairs of neighbor cells; $\bm{c}$ are vectors
	defining the positions of each atom of the lattice basis within the cell.
	Each $\bm{e}_i$ vector represents a nearest-neighbor bond vector, connecting the
	first atom of a unit cell to the second atom of a cell at position $\bm{f}_i$.

	\begin{table*}
		\begin{center}
			\begin{tabular}{| l | c | c |}
				\hline
				&
				Honeycomb &
				Diamond \\ \hline\hline
				($\bm{a}$) &
				 $\{\sqrt{3}, 0\}$, $\{\frac{\sqrt{3}}{2}, \frac{3}{2}\}$,
					$\{-\frac{\sqrt{3}}{2},\frac{3}{2}\}$ &
				$\frac{1}{2} \{0,1,1\}$, $\frac{1}{2} \{1,0,1\}$, $\frac{1}{2} \{1,1,0\}$,
				$\frac{1}{2} \{0,-1,1\}$, $\frac{1}{2} \{1,0,-1\}$, $\frac{1}{2} \{-1,1,0\}$ \\ \hline
				($\bm{c}$) &
				$\{0,0\}$, $\{0,1\}$ &
				$\{0,0,0\}$, $\frac{1}{4} \{1,1,1\}$ \\ \hline
				($\bm{f}$) &
				$\{0,0\}$, $\{\frac{\sqrt{3}}{2}, -\frac{3}{2}\}$,
					$\{-\frac{\sqrt{3}}{2}, -\frac{3}{2}\}$ &
				$\frac{1}{2}\{0,-1,-1\}$, $\frac{1}{2}\{-1,0,-1\}$, $\frac{1}{2}\{-1,-1,0\}$,
					$\{0,0,0\}$ \\ \hline
				($\bm{e}$) &
				$\{0,1\}$, $\{\frac{\sqrt{3}}{2},-\frac{1}{2}\}$,
					$\{ -\frac{\sqrt{3}}{2}, -\frac{1}{2}\}$ &
				$\frac{1}{\sqrt{3}}\{1,-1,-1\}$, $\frac{1}{\sqrt{3}}\{-1,1,-1\}$,
					$\frac{1}{\sqrt{3}}\{-1,-1,1\}$, $\frac{1}{\sqrt{3}}\{1,1,1\}$, \\ \hline
			\end{tabular}
			\caption{
				Particular forms for the honeycomb and diamond lattice vectors used
				in our calculations.
				\label{tab:lattice}}
		\end{center}
	\end{table*}
		
\begin{figure}[!ht]
		\centering (a) \par\smallskip
		\includegraphics[width=0.8\linewidth]{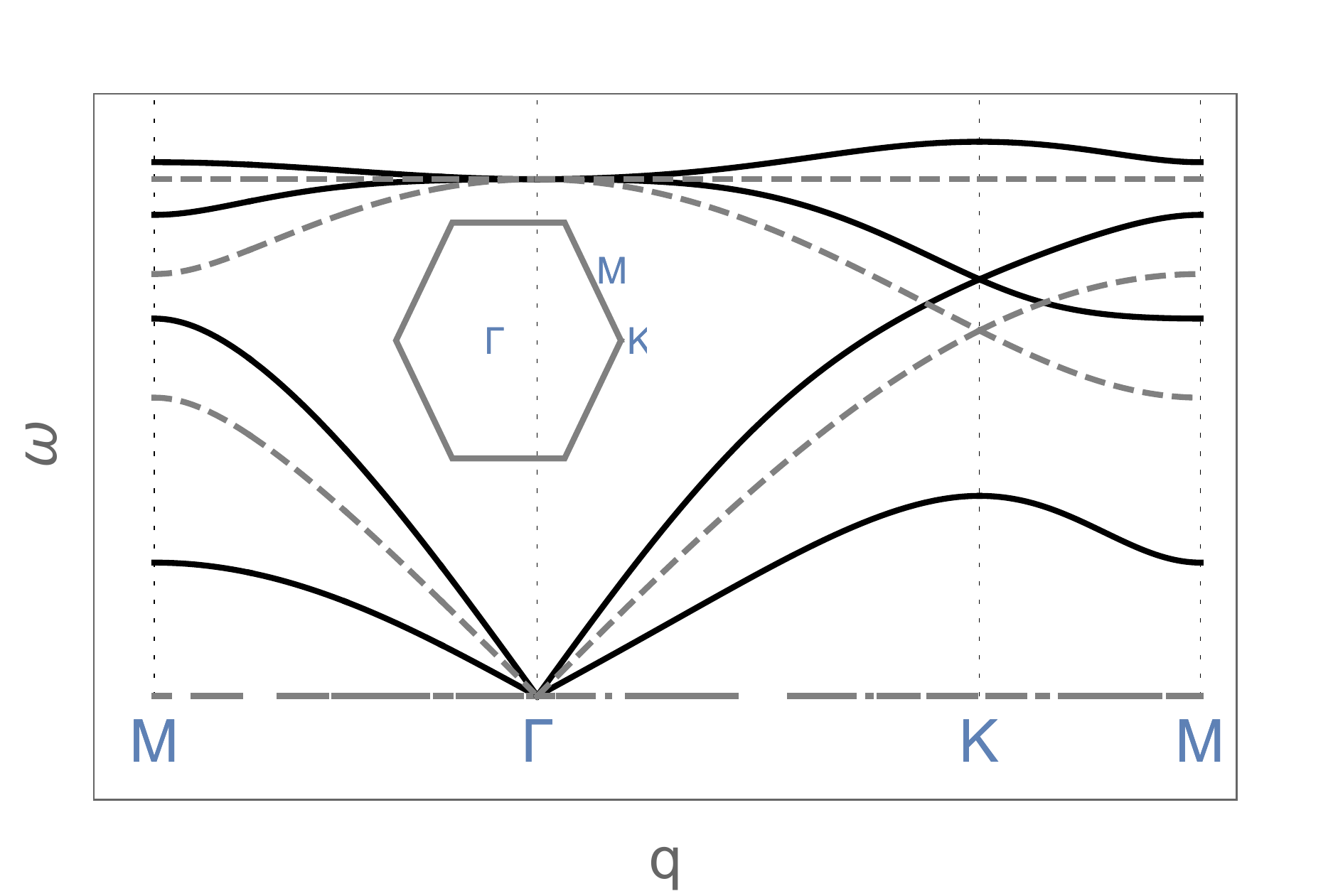}
		
		\centering (b) \par\smallskip
		\includegraphics[width=0.8\linewidth]{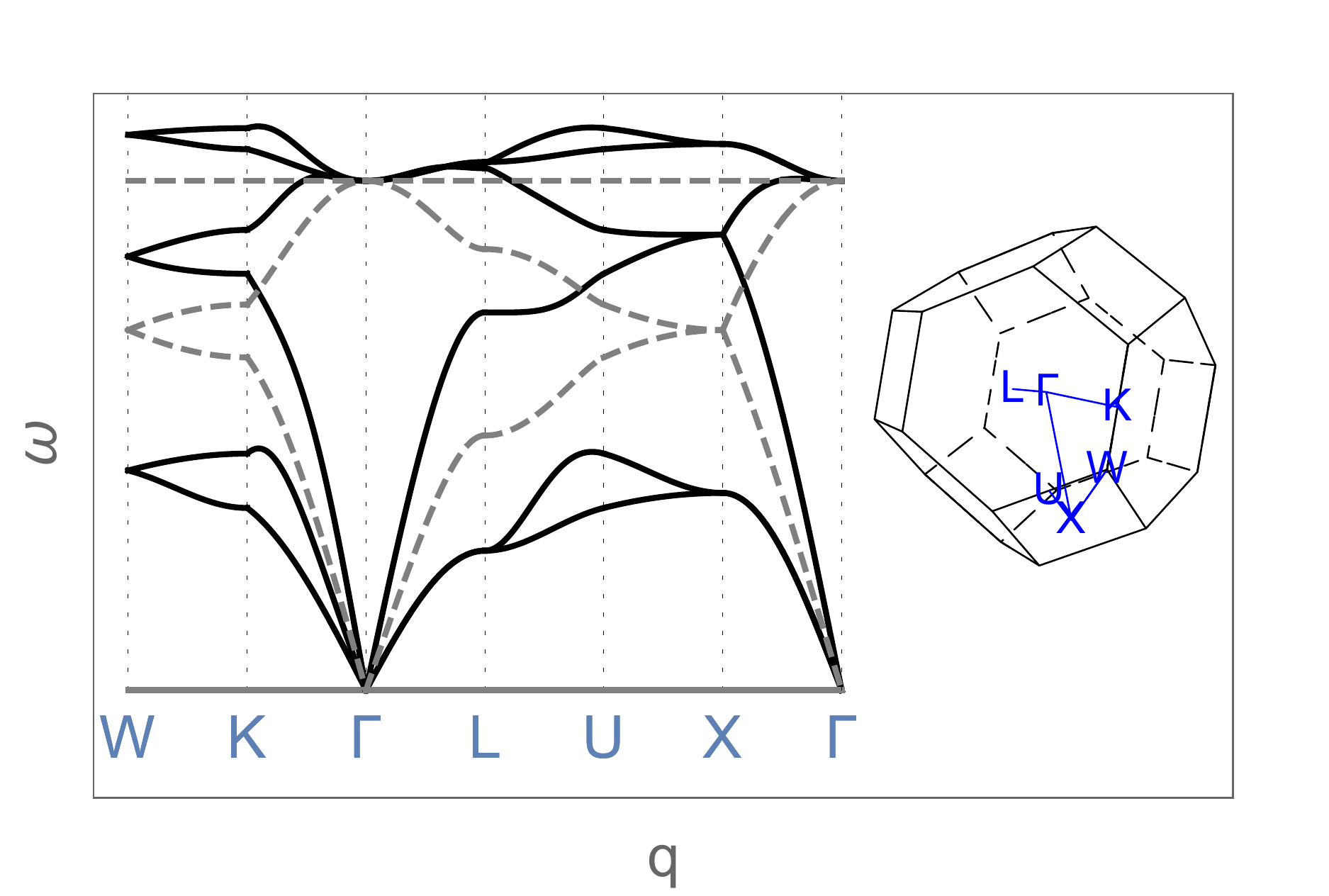}
		\caption{
			Dispersion curves of the honeycomb (a) and diamond (b) lattices,
			with $k_a=1$, and $k_b=k_c=0$ (dashed curves) and $k_b=k_c=0.1$
			(solid curves).
			\label{fig:dispersion}}
	\end{figure}

	The honeycomb lattice has four degrees of freedom per unit cell and thus four
	phonon branches, two of which are optical phonons and two of which are acoustic
	phonons with one branch of compressional phonons that reduce to longitudinal
	phonons in the long wavelength limit and one branch with zero frequency for all
	wavenumbers in the Brillouin zone~\cite{LiarteLub2016}.
	Figure \ref{fig:dispersion}a shows dispersion curves of the homogeneous honeycomb
	lattice with $k_a=1$, $k_b = k_c = 0$ (dashed-grey), and $k_b=k_c=0.1$ (black-solid)
	along symmetry lines $M \Gamma$, $\Gamma K$, and $K M$.
	In the $p_b=p_c$ plane, for $q_x=0$ and $q_y \in [0,2\pi /3]$ ($\Gamma M$ line), the
	floppy branch has frequency $\omega_F = \sqrt{2 \, k_b} \, | \sin (3 q_y / 4) |$, which is
	maximum at $M$, defining a characteristic frequency $\omega^*_M = \sqrt{2\,k_b}$.
	There is also a characteristic frequency associated with the floppy branch energy at
	$K$, $\omega^*_K=\sqrt{9 \, k_b /2}$.

	The diamond lattice has six degrees of freedom per unit cell and thus six phonon
	branches, three of which are optical and three of which are acoustic phonons, the
	latter with one compressional branch and two shear branches with modes of zero
	frequency~\cite{LiarteLub2016}.
	Figure~\ref{fig:dispersion}b shows dispersion curves of the diamond lattice for
	$k_a=1$, $k_b=k_c=0$ (grey-dashed), and $k_b=k_c=0.1$ (solid), along several
	symmetry lines passing through the symmetry points $W$, $K$, $\Gamma$, $L$,
	$U$, and $X$.
	For $p_b=p_c$, we can use the symmetry points $X$ and $L$ to derive two
	vanishing characteristic frequencies associated with the two degenerate lowest
	branches, $\omega_X^{*}=\sqrt{4\, k_b}$, and $\omega_L^*=\sqrt{2\,k_b}$,
	respectively.
	
	The continuum elastic energy of the honeycomb lattice, which has hexagonal symmetry,
	is isotropic, with a bulk modulus $B$ and a single shear modulus $G$:
	\begin{equation}
		\Delta U
			= \frac{B}{2} \left( u_{xx} + u_{yy} \right)^2
				+ 2 \, G \left[ {u_{xy}}^2  + \frac{1}{4}\left( u_{xx} - u_{yy} \right)^2 \right],
	\end{equation}
	where $u_{i j}$ are the components of the linearized strain tensor.
	We can write the elastic moduli in terms of energy coupling constants as
	\begin{equation}
		B
			= \frac{3}{4} k_a + \frac{9}{4} \left(k_b + k_c \right), \quad
		G
			= \frac{9}{8} \, (k_b + k_c).
		\label{eq:honeycomb_moduli}
	\end{equation}
	The diamond lattice, which has cubic symmetry, has the elastic energy:
	\begin{align}
		\Delta U
			& = \frac{1}{2} B \left( u_{xx} + u_{yy} + u_{zz} \right)^2
				+ \frac{1}{3} \, G_1 \left[ \left( u_{xx} - u_{yy} \right)^2
			\right. \nonumber \\ 
			& \quad \left.
				+ \left( u_{xx} - u_{zz} \right)^2 + \left( u_{zz} - u_{yy} \right)^2 \right]
				\nonumber \\
			& \quad + 2 \, G_2 \left( {u_{xy}}^2 + {u_{xz}}^2 + {u_{yz}}^2 \right).
	\end{align}
	The bulk modulus ($B$) and the two shear moduli ($G_1$ and $G_2$)
	can be written in terms of the stretching energy constants as
	\begin{equation}
		B 
			= \frac{k_a}{12} + \frac{k_b+k_c}{3}, \quad
		G_1
			= \frac{k_b+k_c}{8}, \quad
		G_2
			= \frac{k_b+k_c}{4}.
		\label{eq:diamond_moduli}
	\end{equation}
	
\section{Effective medium theory}
	\label{sec:cpa}
	
	In this section, we provide additional details of our effective-medium theory.
	In Section~\ref{subsec:equations}, we present a brief derivation of the set of
	self-consistent equations for the effective-medium energy constants $k_a$, $k_b,$
	and $k_c$.
	In Section~\ref{subsec:analytical}, we perform some explicit calculations for particular
	cases of interest in the vicinity of the jamming ($J$) and the crossover ($X$) points.
	In Section~\ref{subsec:dos}, we calculate the density of states near the jamming
	and rigidity percolation transitions, and explore the crossover from elastic to isostatic
	behavior.
		
	\subsection{Derivation of the EMT self-consistent equations}
		\label{subsec:equations}
		
		We implement an adaptation of the Coherent Potential
		Approximation~\cite{feng85, mao11, mao13a, mao13b} by replacing the
		disordered network by a homogeneous lattice with adequately defined 
		\emph{effective-medium} energy constants $k_\alpha$, with
		$\alpha$ representing one of the three sub-lattices $\{a,b,c\}$.
		To determine $k_\alpha$, we start with a homogeneous system with unit energy
		constants for all bonds in the lattice. 
		We then single out one bond, and assign to it a random energy constant
		$k_\alpha^\prime$ satisfying the bimodal probability distribution
		\begin{equation}
			P(k_\alpha^\prime)
				= p_\alpha \delta (k_\alpha^\prime - 1) + (1-p_\alpha) \delta
					(k_\alpha^\prime).
		\end{equation}
		We treat this bond replacement as a perturbation of the dynamical matrix,
		\begin{equation}
			D^V ({\bm{q}, \bm{q}^\prime} )
				= D ({\bm{q}, \bm{q}^\prime}) + V ({\bm{q}, \bm{q}^\prime}),
		\end{equation}
		with
		\begin{equation}
			V ({\bm{q}, \bm{q}^\prime})
				= ( k_\alpha^\prime-k_\alpha ) \bm{B}^\alpha_{1} ({\bm{q}})
					\bm{B}^\alpha_{1} {(-\bm{q}^\prime}),
		\end{equation}
		where $\bm{B}^\alpha_{1} ({\bm{q}})$ is given by Eqs.~(\ref{eq:Ba}-\ref{eq:Bc}).
		We then use the perturbed ($\mathcal{G}^V$) and unperturbed
		($\mathcal{G}$) retarded Green's functions to define the $T$
		matrix~\cite{elliott74},
		\begin{equation}
			\mathcal{G}^V ({\bm{q}, \bm{q}^{\prime}} )
				= N_c \, \delta_{\bm{q},\bm{q}^\prime} \, \mathcal{G} ({\bm{q}} )
					+ \mathcal{G} ({\bm{q}} )
					\cdot T ({\bm{q}, \bm{q}^{\prime}} ) \cdot
					\mathcal{G} ({\bm{q}^{\prime}} ),
		\end{equation}
		where,
		\begin{equation}
			T ({\bm{q}, \bm{q}^\prime} )
				= V ({\bm{q}, \bm{q}^\prime})
					\left[\frac{1}{1 + (k_\alpha^\prime / k_\alpha- 1)
					h_\alpha(\omega) }\right],
		\end{equation}
		with
		\begin{equation}
			h_\alpha(\omega)
				= \frac{k_\alpha}{\tilde{z}_\alpha N_c}\sum_{\bm{q}} \, \text{Tr} \,
					K_{\alpha}({\bm{q}}) \cdot \mathcal{G}({\bm{q}},\omega),
			\label{eq:h_definition}
		\end{equation}
		where $K_\alpha(\qv)$ is the normalized stiffness matrix, Tr denotes a trace over the $m D$-dimensional matrix
		$K \cdot \mathcal{G}$, with $m$ denoting the number of sites per unit cell.
		The unperturbed Green's function can be written as
		\begin{equation}
			\mathcal{G}({\bm{q}}, \omega)
				= \left[ D({\bm{q}}) - w(\omega) I \right]^{-1},
		\end{equation}
		where $D(\qv) = \sum_\alpha D_\alpha$, $D_\alpha = k_\alpha k_\alpha$, $I$ is an $m D$-dimensional identity matrix, and
		$w(\omega) = \omega^2 +i\gamma \omega$, with $\gamma$ denoting the
		drag coefficient.
		See references~\cite{mao11,mao13a} for more details on the derivation of
		$T$.
		Note that $h_\alpha$ satisfies the sum rule
		\begin{equation}
			\sum_\alpha \tilde{z}_\alpha h_\alpha (\omega)
				= m D \left[1 + \frac{w(\omega)}{N_c} \sum_{\bm{q}} \text{Tr} \,
					\mathcal{G} ({\bm{q}}, \omega) \right].
			\label{eq:sum_rule}
		\end{equation}
		Using the effective-medium approximation
		$\langle T_{\bm{q}, \bm{q}^\prime} \rangle = 0$, with $\langle \cdot \rangle$
		denoting an average with respect to the probability distribution
		$P(k^\prime_\alpha)$, we find the set of self-consistent equations that
		determine $k_a$, $k_b$ and $k_c$.
		\begin{equation}
			k_\alpha
				= \frac{p_\alpha - h_\alpha (\omega)}{1- h_\alpha(\omega)}, \quad
					\text{for } \alpha \in \{a,b,c\}.
			\label{eq:EMT_equations}
		\end{equation}
		Figure~\ref{fig:plots} shows selected simulation and EMT results for the moduli
		of the HLM.
		We will discuss details of the simulation setup in Section~\ref{sec:simulation}.

		\begin{figure}[!ht]
			\begin{minipage}[t]{0.48\linewidth}
				\centering (a) \par\smallskip
				\includegraphics[width=\linewidth]{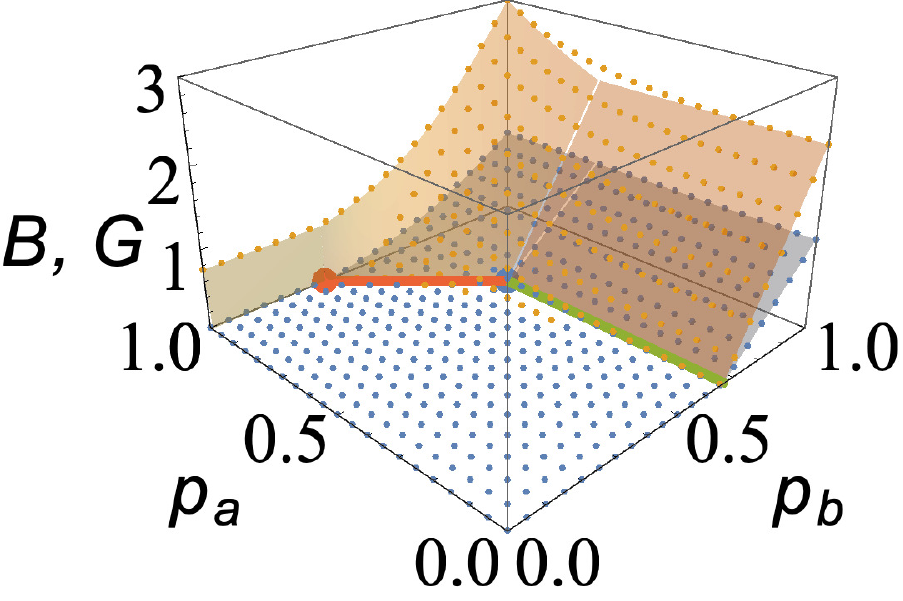}
			\end{minipage}
			\hspace{0.01cm}
			\begin{minipage}[t]{0.48\linewidth}
				\centering (b) \par\smallskip
				\includegraphics[width=\linewidth]{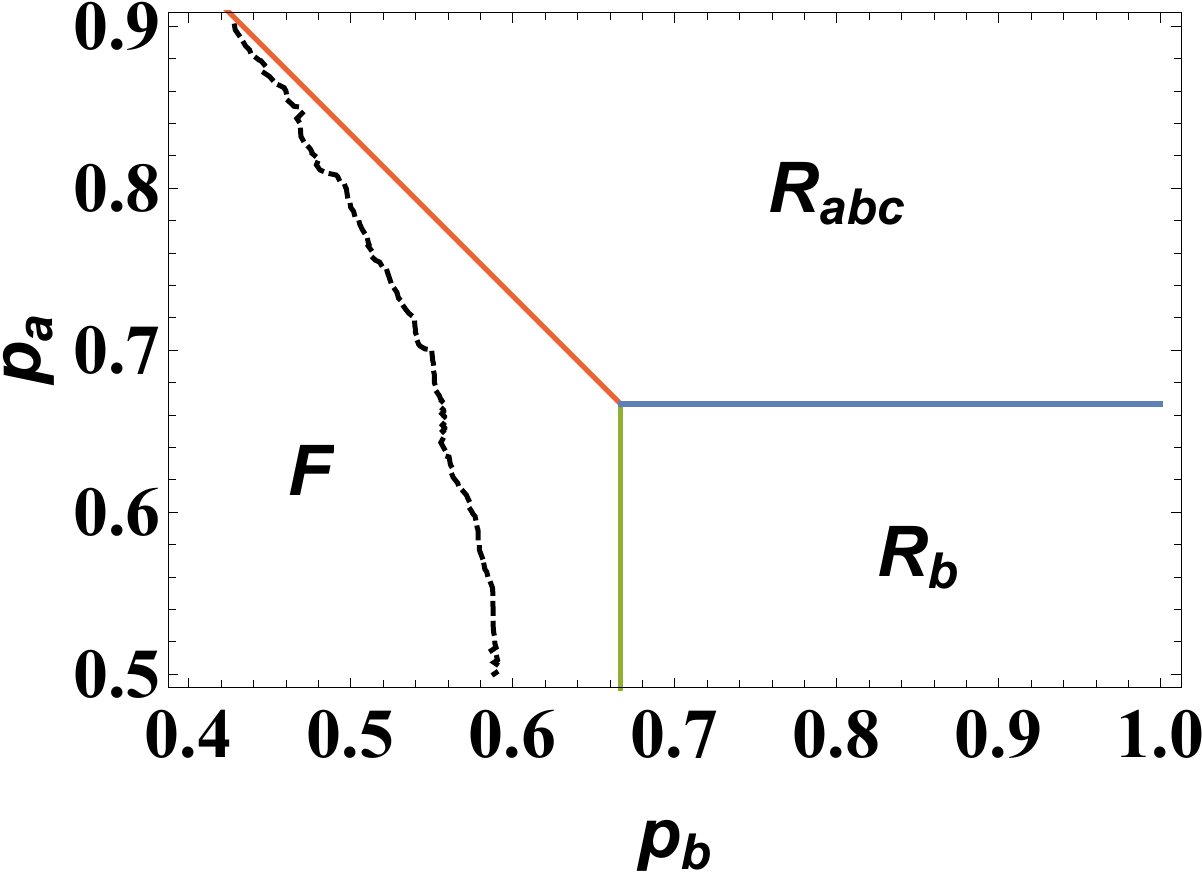}
			\end{minipage}
			
			\vspace{0.1cm}
			\begin{minipage}[t]{0.48\linewidth}
				\centering (c) \par\smallskip
				\includegraphics[width=\linewidth]{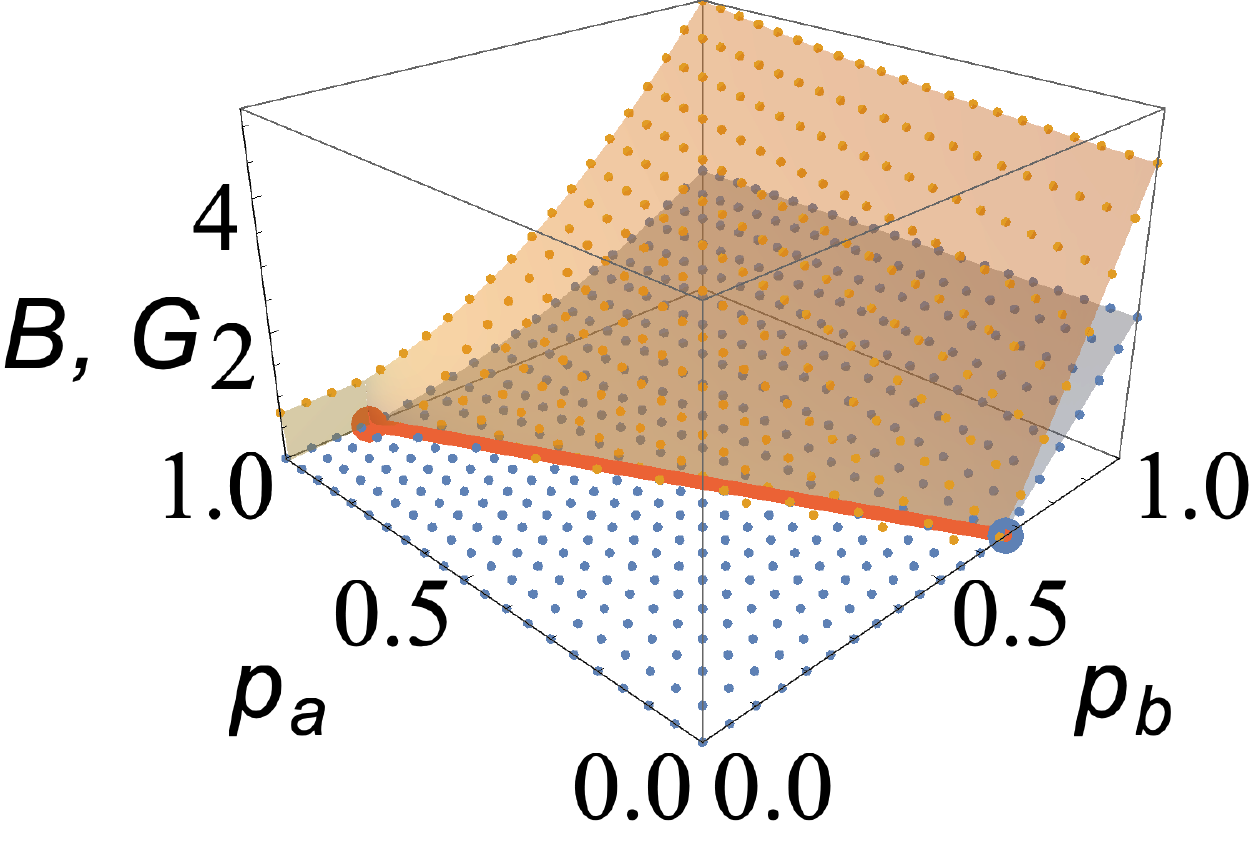}
			\end{minipage}
			\hspace{0.01cm}
			\begin{minipage}[t]{0.48\linewidth}
				\centering (d) \par\smallskip
				\includegraphics[width=\linewidth]{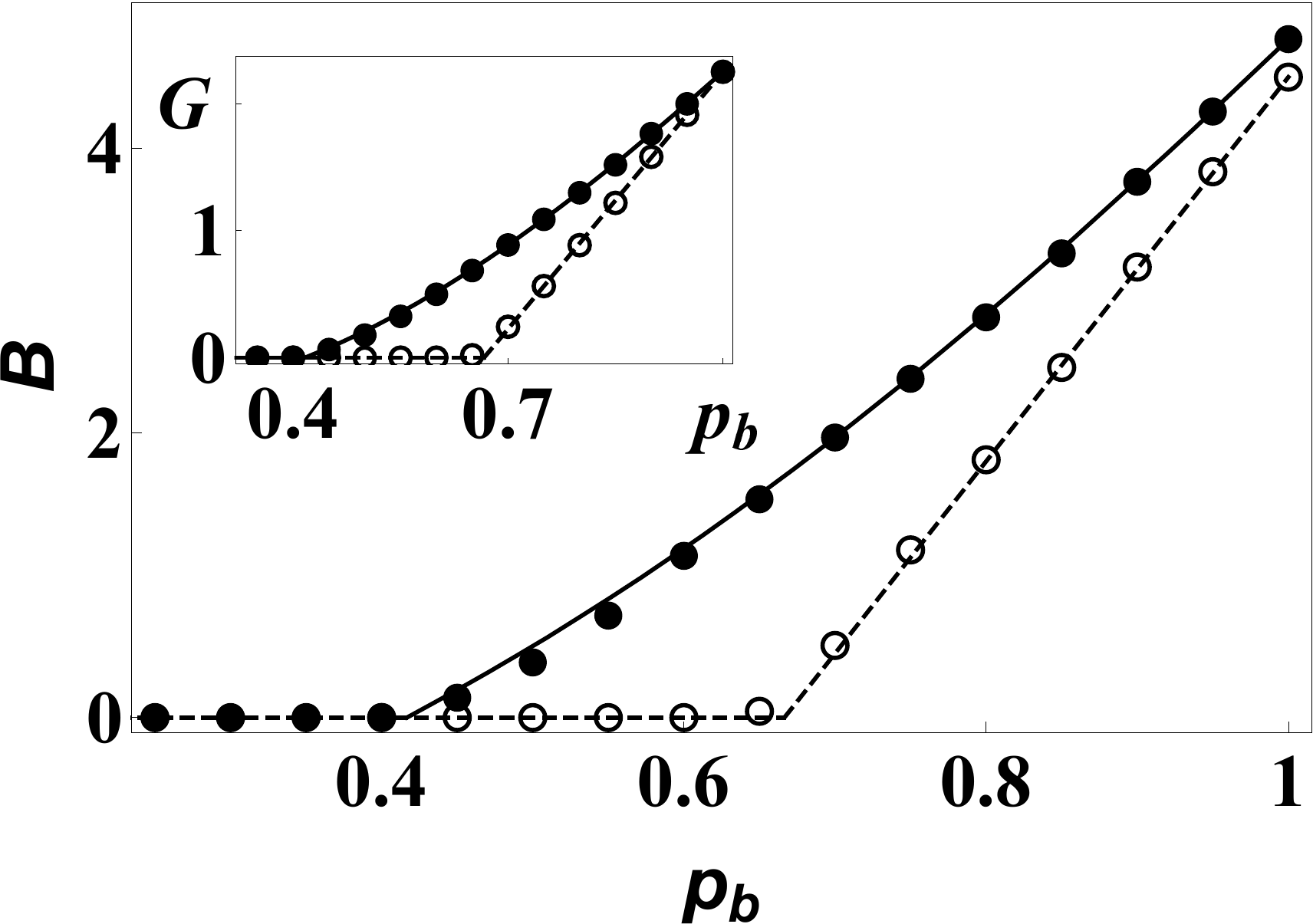}
			\end{minipage}
			\caption{(a) and (c): Simulations (points) and EMT solutions (surfaces) for the
			bulk (yellow) and shear (blue) moduli as a function of $p_a$ and $p_b$ for
			(a) $p_c=0$ and (c) $p_c=p_b$.
			The red lines correspond to $JX$ and $JY$ on Fig~1e of the main text,
			respectively, and the green line to $XY$ on the same figure.
			(b) Phase diagram in $p_a \times p_b$ space showing floppy ($F$), and rigid
			phases ($R_{abc}$ and $R_b$).
			The black dashed line corresponds to approximated rigidity percolation
			thresholds according to the simulations.
			The red, green and blue lines correspond to lines $JX$, $XY$ and $XZ$ in
			Fig.~1e of the main text.
			Note that there is good agreement between simulations and EMT for the
			rigidity percolation threshold near the jamming point, but not near the $X$
			point.
			The jamming point itself does not have redundant springs, so that it is not
			surprising that the EMT works better there than elsewhere.
			Note that the simulations miss the multicritical point where $F$, $R_b$ and
			$R_{abc}$ meet, and exhibit  a smooth transition curve from $F$ to a rigid phase.
			Similarly, the simulations to do show a kink in $G$ or $B$ along the EMT
			$R_b$-$R_{abc}$ phase boundary, likely indicating that fluctuations destroy the
			EMT $R_b$-$R_{abc}$ phase transition.
			(d) Bulk and shear (inset) moduli as a function of $p_b$ for $p_c=p_b$,
			$p_a=0.5$ (filled circles and solid curves) and $p_a=0$ (open circles and dashed
			curves).
			\label{fig:plots}}
		\end{figure}	
	
	\subsection{Approximate solutions for particular cases}
		\label{subsec:analytical}
			Here we present details of our EMT calculations for two particular cases of
			the HLM:
			(1) near the jamming point $J$ in the $p_b=p_c$ plane
			(section~\ref{subsubsec:nearJ});
			(2) near the crossover point $X$ in the $p_c=0$ plane
			(section~\ref{subsubsec:nearX}).
			Although we show results for these particular cases, our final results are
			applicable in general near the jamming and crossover lines of both the
			HLM and DLM.

		\subsubsection{Solutions near the jamming point}
			\label{subsubsec:nearJ}

			Here we derive approximate solutions near the jamming line of the HLM.
			We perform calculations in the $p_b=p_c$ plane, though our results are
			general and applicable near the jamming line of both honeycomb and
			diamond lattices.
			For convenience, let $p_2 \equiv p_b = p_c$, $k_2 \equiv k_b=k_c$,
			$\tilde{z}_2 \equiv \tilde{z}_b+\tilde{z}_c$, and
			$D_{2} ({\bm{q}}) \equiv D_{b} ({\bm{q}}) + D_{c} ({\bm{q}})=k_2 K_2(\qv)$, so that
			\begin{equation}
				k_a
					= \frac{p_a - h_a (\omega)}{1- h_a(\omega)}, \quad
				k_2
					= \frac{p_2 - h_2 (\omega)}{1- h_2(\omega)},
				\label{eq:EMT_equations_2}
			\end{equation}
			where,
			\begin{equation}
				h_2 (\omega)
					= \frac{k_2}{(\tilde{z}_b + \tilde{z}_c)\,N_c} \sum_{\bm{q}} \,
						\text{Tr} \, K_{2} ({\bm{q}}) \cdot \mathcal{G}({\bm{q}},\omega).
				\label{eq:h_definition2}
			\end{equation}

\begin{figure}[!ht]
				\centering
				\includegraphics[width=0.8\linewidth]{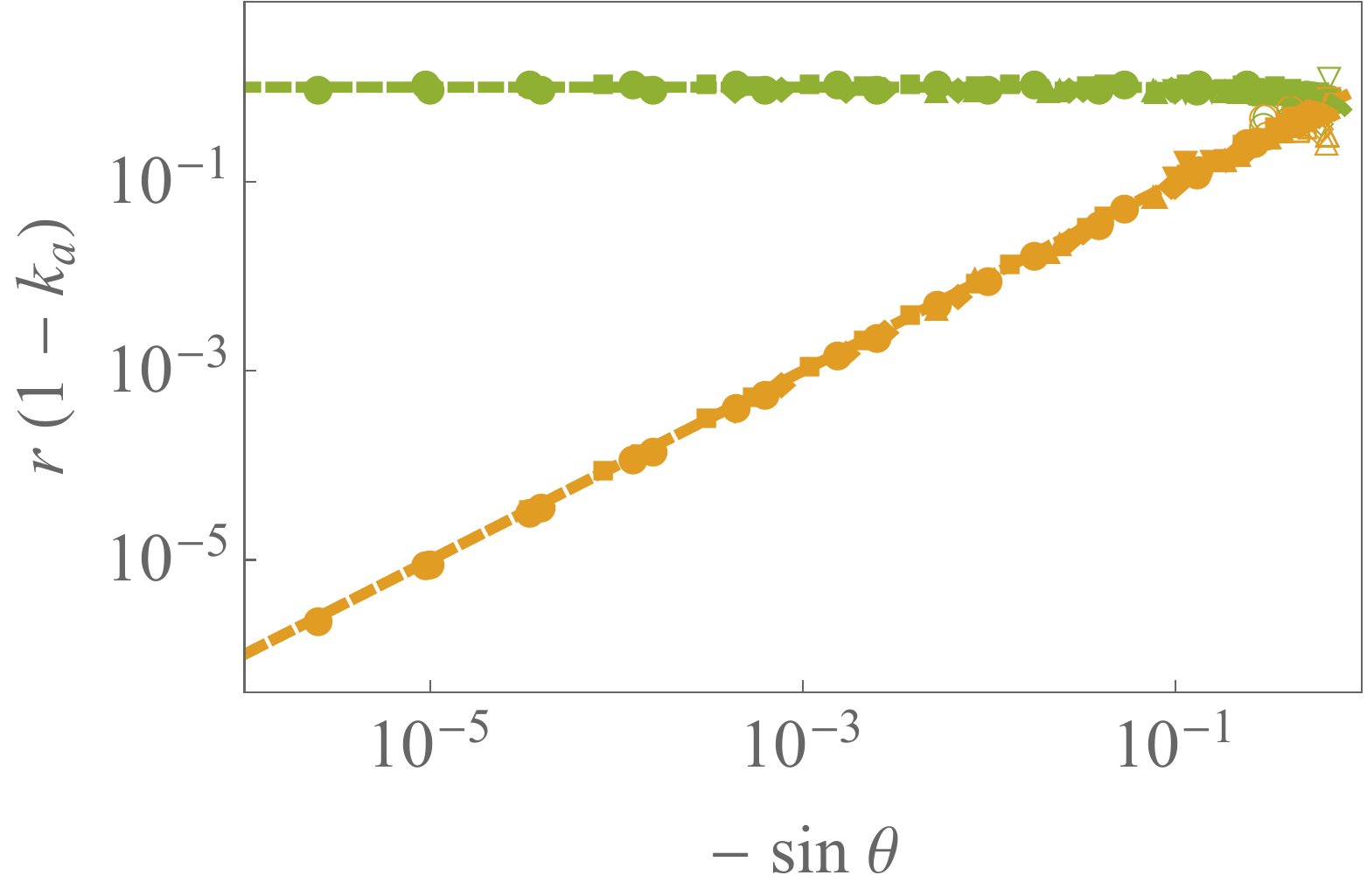}
				\caption{%
					Rescaled numerical solutions of the dynamical CPA equations near
					jamming, in the over-damped limit $w(\omega) = i\gamma \omega$,
					with $r \, e^{i \theta} \equiv 1+c_J \, k_2 / \Delta p_a$.
					Green and yellow symbols represent the real and imaginary part of
					$r \, (1-k_a)$, respectively, obtained from the solutions of the dynamic-CPA
					equations for paths approaching the jamming point from the rigid (through
					the $p_a = 7/6 - p_b$ line with $\Delta \tilde{p} > 0$) and non-rigid (through
					the $p_b =1/6$ line with $\Delta \tilde{p} <0$) phases.
					For each plot, we show several values of $\omega \in (10^{-8},10^{-2}]$ and
					$\delta \in (10^{-6},10^{-1}]$, where
					$\delta = \sqrt{(\Delta p_a^J)^2+(\Delta p_b^J)^2}$ is the distance to the
					jamming point in $p_a \times p_b$ space.
					The green and yellow dashed lines correspond to our approximated
					analytical relation given by Eq.~\eqref{eq:ka-kb-scaling_nearJ}.
					\label{fig:ka-kb-scaling_nearJ}}
			\end{figure}

			At small non-zero frequency ($0 < w / k_2 \ll 1$), near the jamming point,
			$k_2 \ll 1$, and direct numerical calculations give
			\begin{equation}
				h_a
					\approx 1- c_J \frac{k_2}{k_a} + v_a \frac{w}{k_a},
				\quad
				h_2
					\approx p_2^J + c_J\, \nu \, \frac{k_2}{k_a} + v_2 \frac{w}{k_2},
			\end{equation}
			where $p_2^J \equiv (mD-\tilde{z}_a)/\tilde{z}_2$,
			$\nu = \tilde{z}_a / \tilde{z}_2$, and our numerical estimates give
			$c_J \approx 3$, $v_a\approx 2.2$ and $v_2 \approx 0.36$ for the HLM.
			Note that the first two terms of the r.h.s. of the equation for $h_2$ are
			chosen so that $h_a$ and $h_2$ are consistent with
			Eq.~\eqref{eq:sum_rule} at zero frequency (the sum rule).
			The approximated CPA self-consistent equations now read
			\begin{align}
				& k_a
					\approx \frac{-\Delta p_a +c_J \, k_2/k_a - v_a w/k_a}{c_J \,
					k_2/k_a - v_a w/k_a},
				\label{eq:ka_cpa_nearJ} \\
				& k_2
					\approx \frac{\Delta p_2 -c_J \, \nu \, k_2/k_a - v_2 w/k_2}{s},
				\label{eq:k2_cpa_nearJ}
			\end{align}
			where $\Delta p_a \equiv 1 - p_a$ and $s \equiv 1 - p_2^J$.
			Dividing \eqref{eq:k2_cpa_nearJ} by \eqref{eq:ka_cpa_nearJ}, and solving
			for $k_2/k_a$ keeping only linear terms in $\Delta p$ and $w$ leads to
			\begin{equation}
				c_J \, \frac{k_2}{k_a}
					= \frac{c_J \, \Delta p_2 + s\, \Delta p_a -c_J \, v_2 w / k_2}{s
					+ c_J \, \nu}.
					\label{eq:eta_cpa_nearJ}
			\end{equation}
			Plugging \eqref{eq:eta_cpa_nearJ} back into \eqref{eq:k2_cpa_nearJ} results
			in a quadratic equation,
			\begin{equation}
				(s+c_J\,\nu){k_2}^2 - \Delta \tilde{p} \, k_2 + v_2 w
					= 0,
				\label{eq:sqEqK2}
			\end{equation}
			where $\Delta \tilde{p} = \Delta p_2-\nu \Delta p_a$, and
			$\Delta p_2 = p_2 - p_2^J$.
			Equation~\eqref{eq:sqEqK2} has the physical solution
			\begin{equation}
				k_2
					= \frac{1}{2(s+\nu \, c_J)} \left[\Delta \tilde{p}
						+ \sqrt{|\Delta \tilde{p}|^2-4(s+\nu c_J) \, v_2 \, w (\omega)}\right].
				\label{eq:dyn_k2_nearJ}
			\end{equation}
			Note that $\text{Re}(k_2) \approx -\text{Im}(k_2)$ at large frequency for
			$w = i \gamma \omega$, since in this case $k_2 \propto \sqrt{i} \propto 1-i$.
			Now we can use Eq.~\eqref{eq:ka_cpa_nearJ} to write down the solution for
			$k_a$ in terms of $k_2$ and $w$:
			\begin{equation}
				k_a
					= \frac{c_J\, k_2-v_a w}{c_J\, k_2-v_a w + \Delta p_a}
					\approx \frac{c_J\, k_2}{c_J\, k_2 + \Delta p_a}.
				\label{eq:ka_nearJ_loww}
			\end{equation}
			This relation is valid at low frequency for paths approaching the jamming point
			from both the floppy and rigid regions in the phase diagram.
			Equation~\eqref{eq:ka_nearJ_loww} can be rewritten as
			\begin{equation}
				1 - k_a
					= \frac{1}{1+c_J \, k_2 / \Delta p_a}.
			\end{equation}
			Let $r$ and $\theta$ be the magnitude and phase of
			$1+c_J \, k_2 / \Delta p_a \equiv r \, e^{i \theta}$.
			Thus,
			\begin{equation}
				r \, (1 - k_a )
					= e^{-i\theta},
				\label{eq:ka-kb-scaling_nearJ}
			\end{equation}
			which is consistent with our numerical solutions of the CPA equations (see
			FIG.~\ref{fig:ka-kb-scaling_nearJ}).
			In the limit of very low frequencies,
			\begin{equation}
				k_2
					\approx
						\displaystyle\frac{[\Delta \tilde{p}]}{s+\nu \, c_J}
								- \displaystyle\frac{v_2 w}{|\Delta \tilde{p}|},
			\end{equation}
			where $[\phi] \equiv (\phi + |\phi|)/2 = 0$ and $\phi$, for $\phi<0$ and $\phi>0$,
			respectively. 
			Thus, for $\Delta \tilde{p}<0$,
			\begin{equation}
				k_a
					\approx \left(1+\frac{\Delta p_a}{c_J k_2} \right)^{-1}
					\approx \left(1-\frac{\Delta p_a |\Delta \tilde{p}|}{c_J \, v_2 \, w} \right)^{-1}
					\approx -\frac{c_J \, v_2 \, w}{\Delta p_a |\Delta \tilde{p}|}.
				\label{eq:ka_nearJ_very-loww}
			\end{equation}
			
			\begin{figure*}[!ht]
				\begin{minipage}[t]{0.48\linewidth}
					\centering (a) \par\smallskip
					\includegraphics[width=0.8\linewidth]{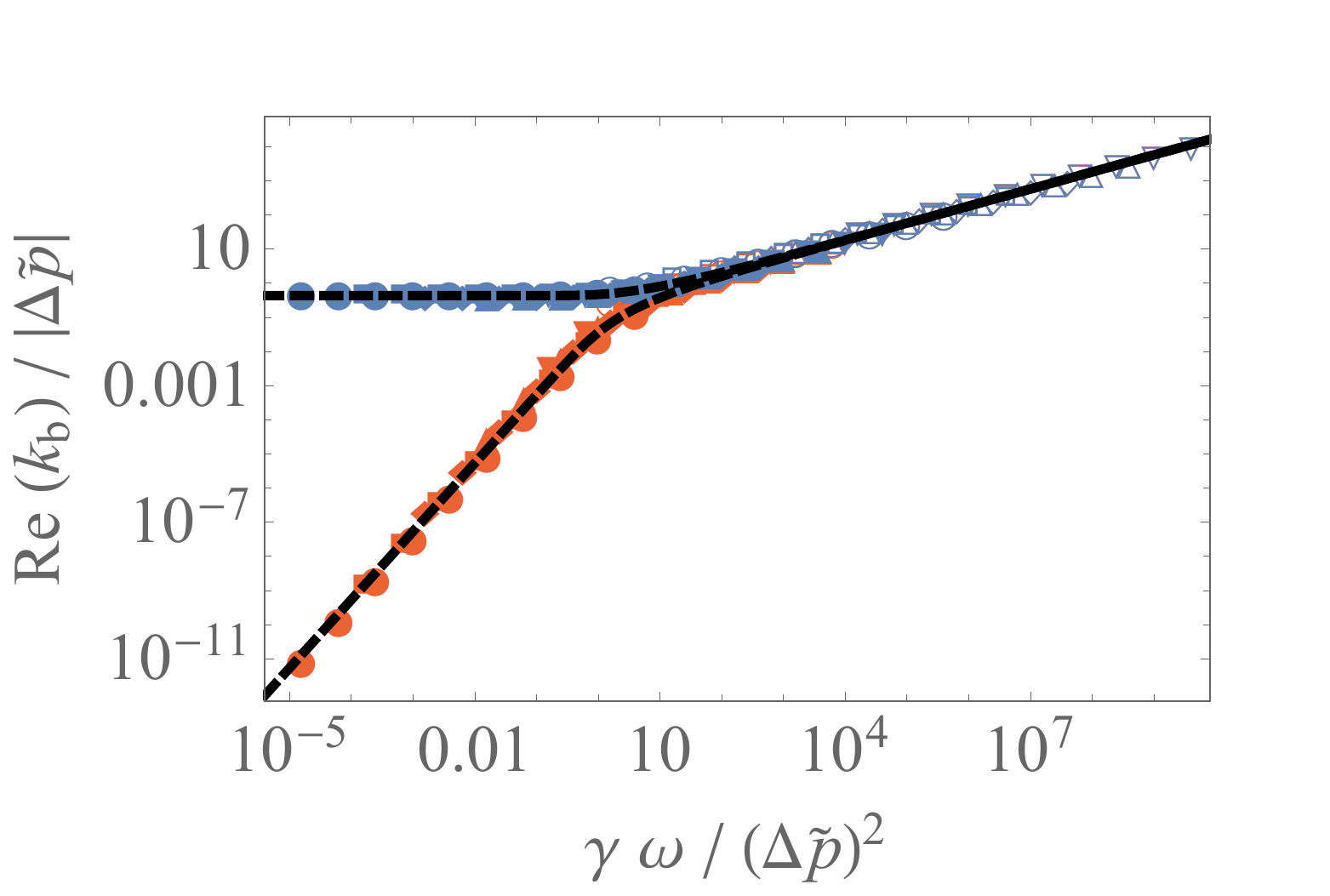}
				\end{minipage}
				\hspace{0.01cm}
				\begin{minipage}[t]{0.48\linewidth}
					\centering (b) \par\smallskip
					\includegraphics[width=0.8\linewidth]{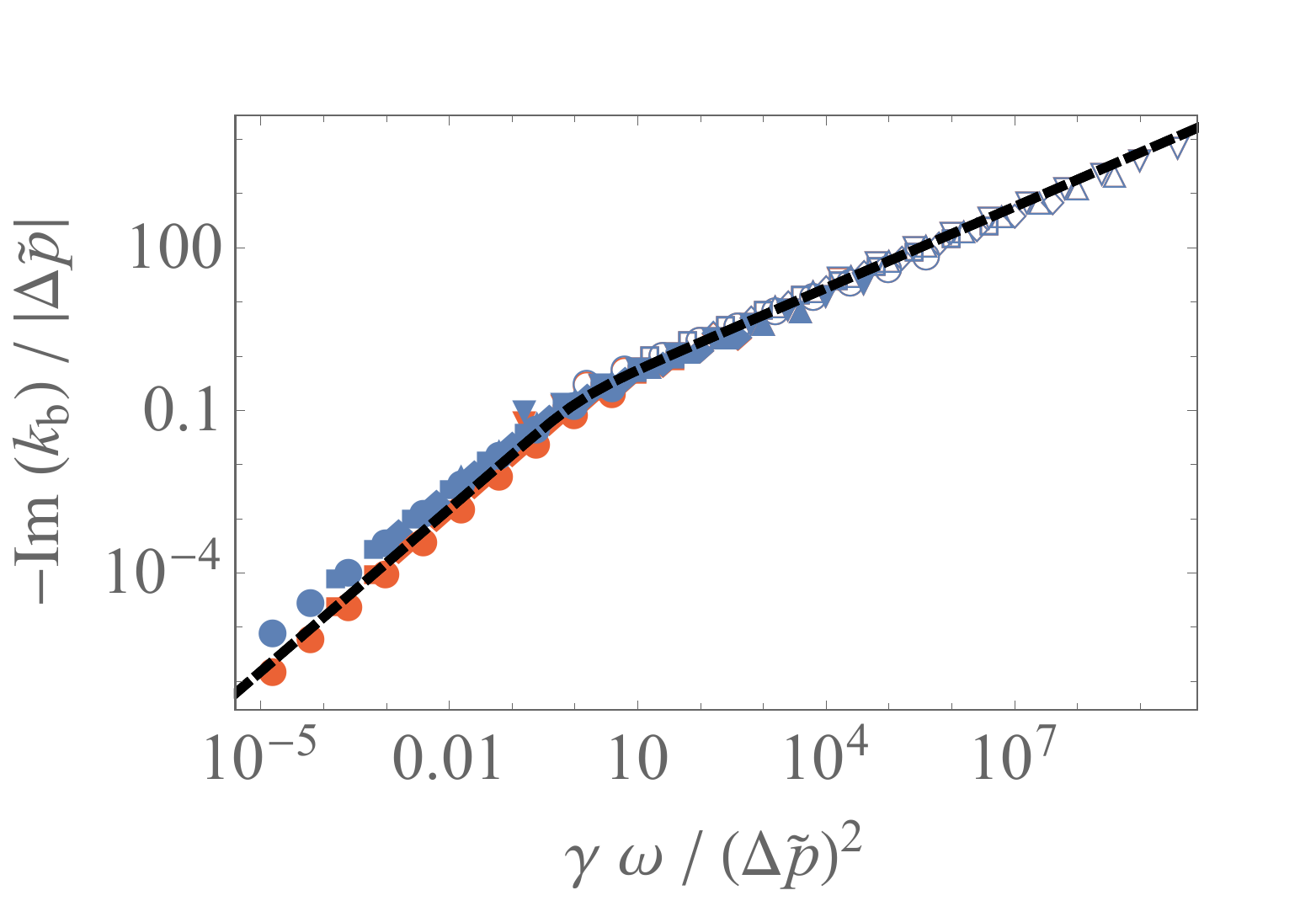}
				\end{minipage}
				\vspace{0.01cm}
				\begin{minipage}[t]{0.48\linewidth}
					\centering (c) \par\smallskip
					\includegraphics[width=0.8\linewidth]{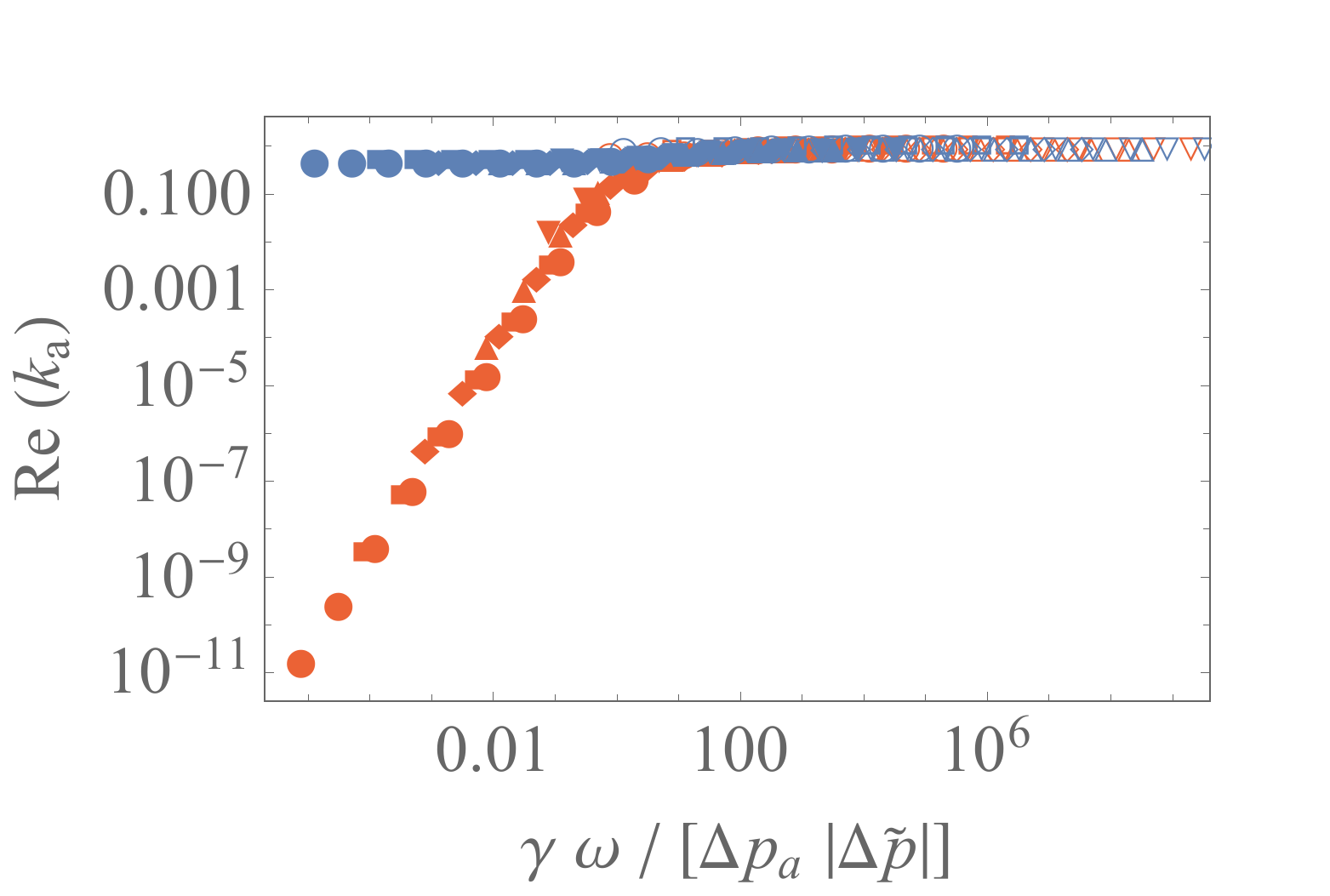}
				\end{minipage}
				\hspace{0.01cm}
				\begin{minipage}[t]{0.48\linewidth}
					\centering (d) \par\smallskip
					\includegraphics[width=0.8\linewidth]{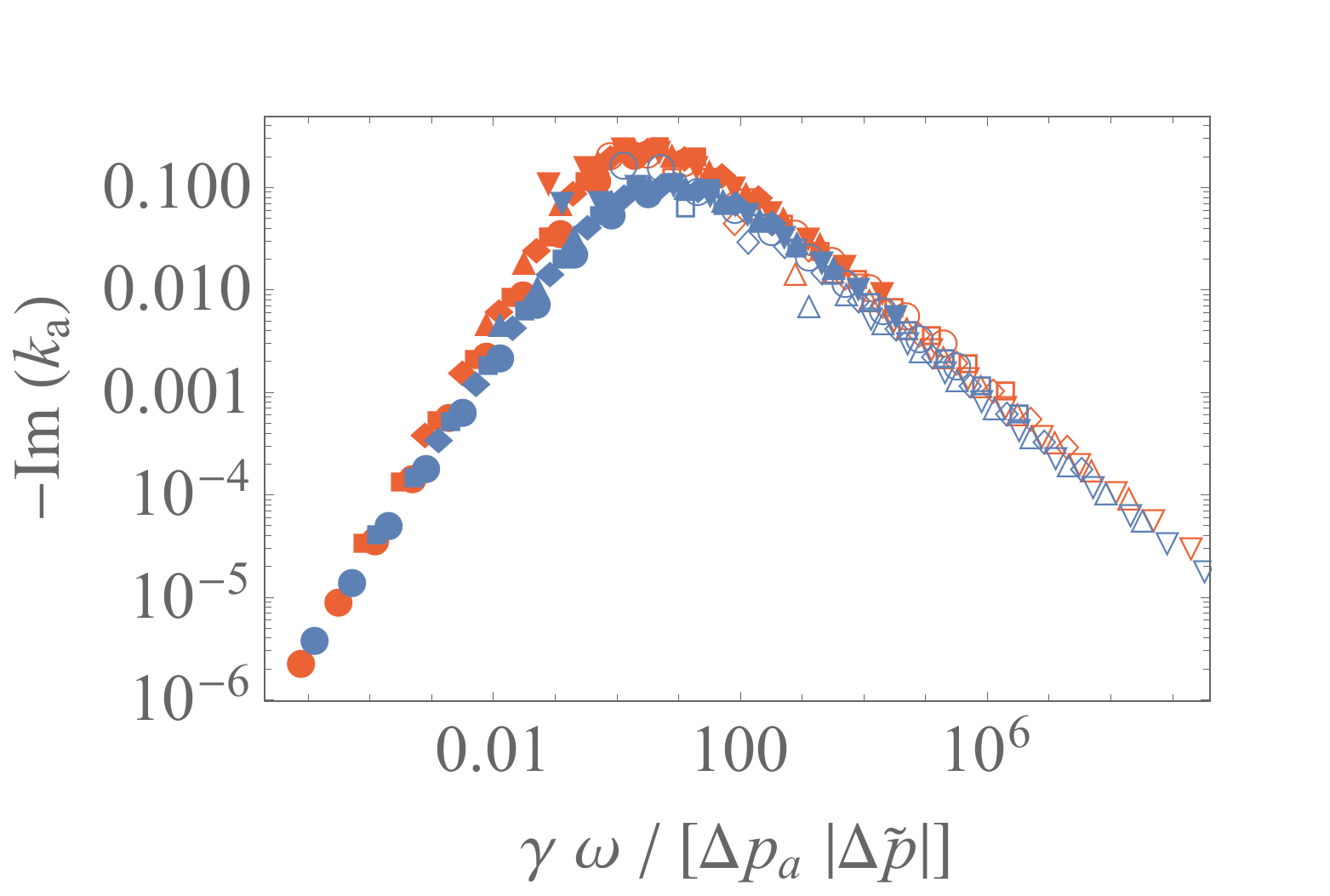}
				\end{minipage}
				\caption{
					Scaling behavior of $k_b$ (a and b) and $k_a$ (c and d) near the jamming
					point for $p_b=p_c$.
					Blue and red symbols represent solutions of the dynamic-CPA equations for
					paths approaching the jamming point from the rigid (through the
					$p_a = 7/6 - p_b$ line with $\Delta \tilde{p} > 0$) and non-rigid (through the
					$p_b =1/6$ line with $\Delta \tilde{p} <0$) phases, respectively.
					For each plot, we show several values of $\omega \in (10^{-8},10^{-2}]$ and
					$\delta \in (10^{-6},10^{-1}]$, where $\delta$ is the distance to the jamming
					point in $p_a \times p_b$ space.
					The black dashed lines on (a) and (b) correspond to our approximated
					analytical solution given by Eq.~\eqref{eq:dyn_k2_nearJ}.
					\label{fig:rheology_nearJ}}
			\end{figure*}

			Figure~\ref{fig:rheology_nearJ} shows scaling plots of $k_b$ (a and b) and $k_a$
			(c and d) for the honeycomb lattice, in the over-damped limit
			$w(\omega) = i\gamma \omega$, and for paths approaching the jamming point at
			constant $p_b=p_c$.
			Note that the bulk viscosity diverges as ${\Delta p_a}^{-2}$ as the jamming point is
			approached from the non-rigid phase (in contrast with $\propto {\Delta p_a}^{-1}$
			near rigidity percolation), which follows from our asymptotic
			solutions~\eqref{eq:dyn_k2_nearJ} and~\eqref{eq:ka_nearJ_loww}.
			Note also that Eq.~\eqref{eq:dyn_k2_nearJ} yields the following scaling behavior
			for the shear modulus:
			\begin{equation}
				G
					\propto |\Delta \tilde{p}| \, \mathcal{S}(w / |\Delta \tilde{p}|^2),
			\end{equation}
			which is consistent with results for soft sphere packings near
			jamming~\cite{tighe11}.
			Finally, our scaling variable for $k_a$ is motivated	by the low-frequency limit of
			the CPA solution given by Eq.~\eqref{eq:ka_nearJ_very-loww}; it is surprising that it
			collapses our numerical solutions even at large
			$\gamma \omega / [\Delta p_a |\Delta \tilde{p}|]$.

		\subsubsection{Solutions near the crossover point}
			\label{subsubsec:nearX}
			
			Here we derive approximate solutions that are valid near the $X$ lines of the
			HLM.
			For mathematical convenience, we consider the $p_c=0$ case, though the
			results are general and applicable near the crossover lines of both the
			honeycomb and diamond lattices.

			At small frequency, near $X$, $k_a, k_b \ll 1$, $k_a / k_b \ll 1$,
			$0 < w / k_a \ll 1$, so that,
			\begin{equation}
				h_a
					\approx p_a^X + c_X \frac{k_a}{k_b} + v_a^X \frac{w}{k_a}, \quad
				h_b
					\approx p_b^X - c_X \nu_X \frac{k_a}{k_b}  + v_b^X \frac{w}{k_b},
			\end{equation}
			where $p_a^X = D / \tilde{z}_a$ and $p_b^X = (m-1)D / \tilde{z}_b$,
			$\nu_X=\tilde{z}_a/\tilde{z}_b$, $c_X \approx 0.1$ for the HLM, and $v_a^X$
			and $v_b^X$ are constants that can be numerically estimated.
			The CPA approximated equations now read
			\begin{align}
				& s_a k_a
					\approx \Delta p_a^X - c_X \frac{k_a}{k_b} - \frac{v_a^X w}{k_a},
					\label{eq:ka_cpa_nearX} \\
				& s_b k_b
					\approx \Delta p_b^X + c_X \nu_X \frac{k_a}{k_b} -
					\frac{v_b^X w}{k_b},
					\label{eq:kb_cpa_nearX}
			\end{align}
			where $\Delta p_a^X = p_a - p_a^X$, $\Delta p_b^X = p_b - p_b^X$,
			$s_a=1-p_a^X$, and $s_b=1-p_b^X$.
			We first multiply~\eqref{eq:ka_cpa_nearX} by $k_a$, and solve the resulting
			quadratic equation for $k_a$,
			\begin{align}
				& \left(s_a +\frac{c_X}{k_b} \right) {k_a}^2 - \Delta p_a^X k_a + v_a^X w
				\nonumber \\ & \quad
					\approx \frac{c_X}{k_b} {k_a}^2 - \Delta p_a^X k_a + v_a^X w
					= 0,
			\end{align}
			where we have ignored $s_a \ll c_X/k_b$.
			Thus,
			\begin{equation}
				k_a
					= \frac{k_b}{2 \, c_X} \left(\Delta p_a^X
						+ \sqrt{(\Delta p_a^X)^2 - \frac{4 \, c_X \, v_a^X w}{k_b}}\right).
			\end{equation}
			For small $w$ (i.e. $4 \, c_X \, v_a^X w/ (k_b (\Delta p_a^X)^2) \ll 1$),
			\begin{equation}
				k_a
					= \frac{k_b}{c_X} [\Delta p_a^X] - \frac{v_a^X w}{|\Delta p_a^X|}.
				\label{eq:low-w_ka_nearX}
			\end{equation}
			Multiplying~\eqref{eq:kb_cpa_nearX} by $k_b$ results in
			\begin{equation}
				s_b {k_b}^2 - \Delta p_b^X k_b - c_X \nu_X k_a + v_b^X w
					\approx 0.
				\label{eq:kb_quadratic_nearX}
			\end{equation}
			Now we use~\eqref{eq:low-w_ka_nearX} in~\eqref{eq:kb_quadratic_nearX}
			to find a low-$w$ solution for $k_b$,
			\begin{align}
				& s_b {k_b}^2 - \Delta p_b^X k_b - c_X \nu_X
					\left(\frac{k_b}{c_X} [\Delta p_{a}^X]
					- \frac{v_a^X w}{|\Delta p_a^X|} \right) + v_b^X w
				\nonumber \\ & \quad
					\approx
					s_b {k_b}^2 - \Delta \tilde{p}_{ab}^X \,k_b
					+ \frac{c_X \nu_Xv_a^X w}{|\Delta p_a^X|}	
					= 0,
			\end{align}
			where
			\begin{equation}
				\Delta \tilde{p}_{ab}^X
					\equiv \Delta p_b^X + \nu_X [\Delta p_{a}^X],
			\end{equation}
			and we have ignored $v_b^X w \ll v_a^X w / |\Delta p_a^X|$.
			Thus,
			\begin{equation}
				k_b
					= \frac{1}{2 s_b} \left(\Delta \tilde{p}_{ab}^X
						+ \sqrt{(\Delta \tilde{p}_{ab}^X)^2
						- \frac{4 s_b c_X \nu_X v_a^X w}{|\Delta p_a^X|}}\right),
			\end{equation}
			which for small $w$ results in,
			\begin{equation}
				k_b
					= \frac{[\Delta \tilde{p}_{ab}^X]}{s_b}
						- \frac{c_X \nu_X v_a^X w}{|\Delta p_a^X||\Delta \tilde{p}_{ab}^X|}.
				\label{eq:low-w_kb_nearX}
			\end{equation}
			Plugging~\eqref{eq:low-w_kb_nearX} back into~\eqref{eq:low-w_ka_nearX} yields
			\begin{equation}
				k_a
					=\frac{[\Delta p_{a}^X] [\Delta \tilde{p}_{ab}^X]}{c_X s_b}
						-\left(1+\frac{\nu_X [\Delta p_{a}^X]}{|\Delta \tilde{p}_{ab}^X|}\right)
						\frac{v_a^X w}{|\Delta p_a^X|}.
			\end{equation}

\begin{figure}[!ht]
				\centering
				\includegraphics[width=0.8\linewidth]{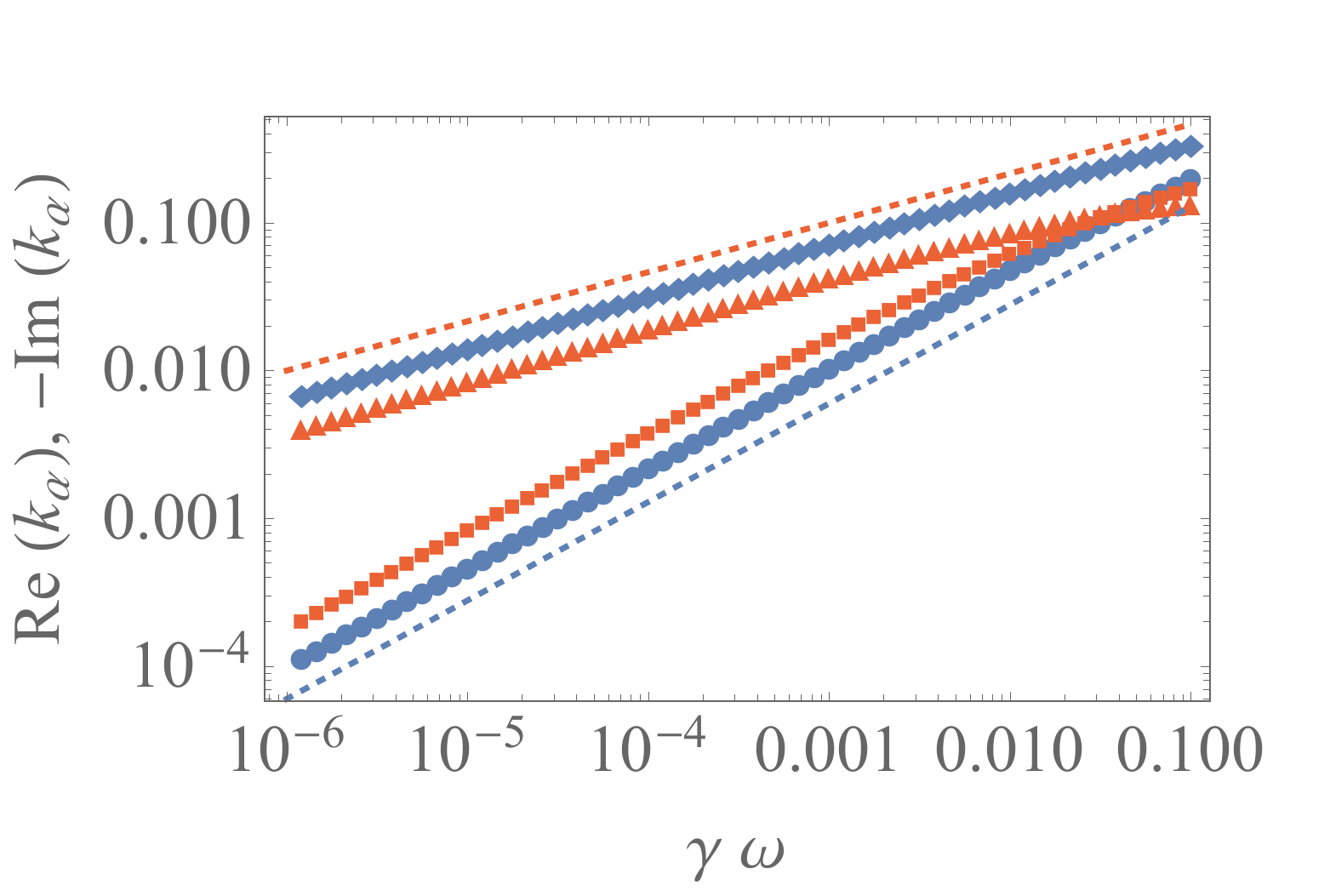}
				\caption{
					Numerical solutions for $k_a$ (circles and squares) and $k_b$
					(diamond and triangles) of the dynamical CPA equations \emph{at the
					$X$ point} of the HLM, for $p_c=0$ and $p_a=p_b=2/3$.
					Blue and red symbols denote the real and minus the imaginary parts
					of $k_\alpha$, with $\alpha \in \{a,b\}$.
					The blue and red dotted lines correspond to power laws associated
					with exponents $2/3$ and $1/3$, respectively.
					\label{fig:viscosities_atX}}
			\end{figure}

			At $X$, $\Delta p_a^X=\Delta p_b^X=0$, so that,
			\begin{align}
				& s_a k_a
					= - c_X \frac{k_a}{k_b} - \frac{v_a^X w}{k_a}
						\Rightarrow \quad
				0
					\approx - c_X \frac{k_a}{k_b} - \frac{v_a^X w}{k_a},
					\label{eq:ka_eq_atX} \\
				& s_b k_b
					= c_X \nu_X \frac{k_a}{k_b} - \frac{v_b^X w}{k_b}
						\Rightarrow \quad
				s_b k_b
					\approx c_X \nu_X \frac{k_a}{k_b},
					\label{eq:kb_eq_atX}
			\end{align}
			where we ignored the terms $s_a k_a \ll c_X k_a/k_b$ and
			$v_b^X w / k_b = (v_b w / k_a) (k_a/k_b) \ll c_X \nu_X (k_a / k_b)$.
			Equations~\eqref{eq:ka_eq_atX} and \eqref{eq:kb_eq_atX} then imply
			$k_a \sim w^{2/3}$ and $k_b \sim w^{1/3}$, which agree with our
			numerical solutions of the CPA equations (see
			FIG.~\ref{fig:viscosities_atX}).
			
	\subsection{Density of states}
		\label{subsec:dos}

		Here we study the small-frequency behavior of the density of states near both the
		jamming and rigidity percolation thresholds.
		We show that the DOS is nearly constant at very small frequencies near the floppy
		phase.
		We describe the characteristic crossover frequency
		$\omega^* \sim \Delta \tilde{p}$ marking the transition from Debye elasticity to
		isostatic behavior.

		\begin{figure*}[!ht]
			\begin{minipage}[t]{0.48\linewidth}
				\centering (a) \par\smallskip
				\includegraphics[width=0.8\linewidth]{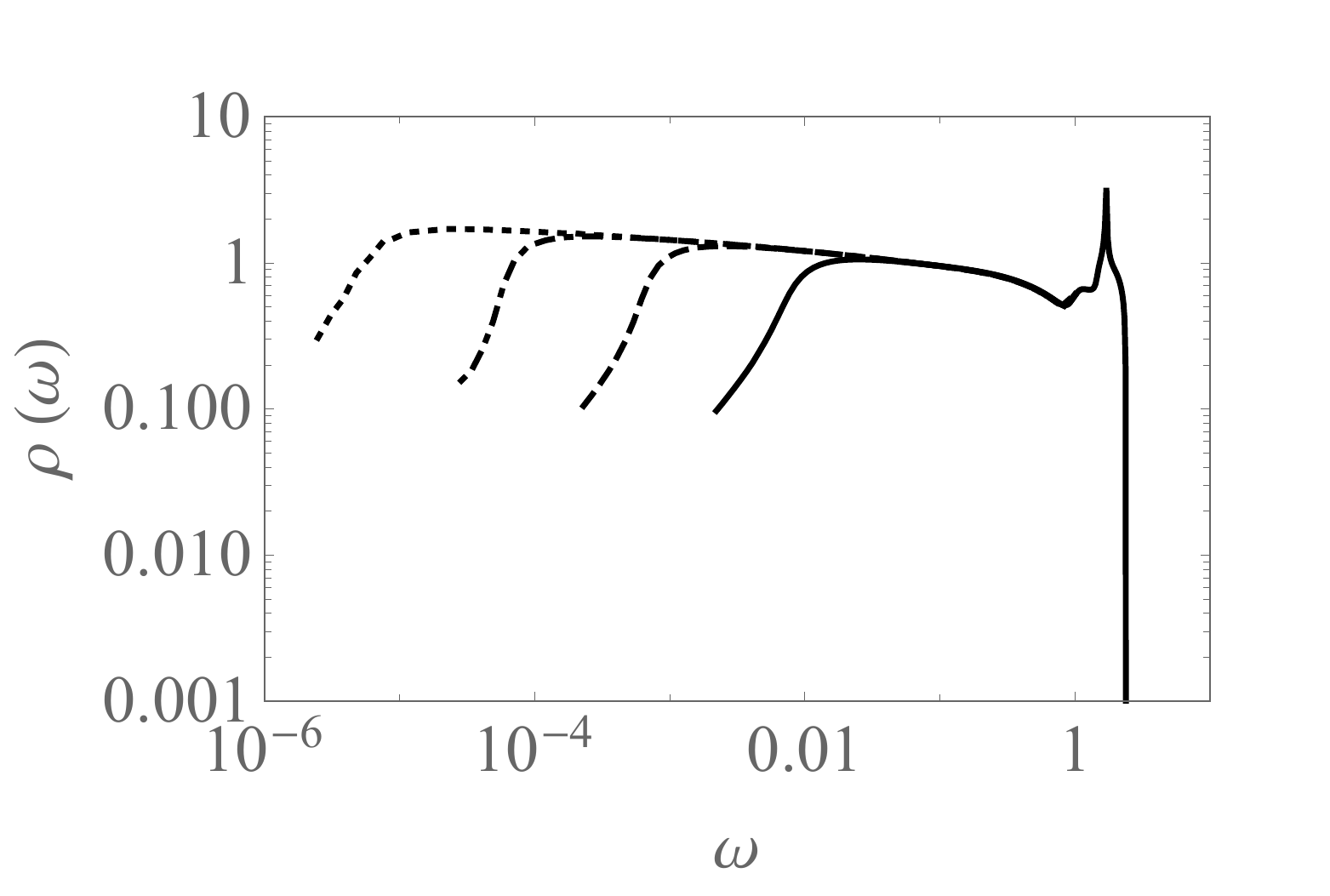}
			\end{minipage}
			\hspace{0.01cm}
			\begin{minipage}[t]{0.48\linewidth}
				\centering (b) \par\smallskip
				\includegraphics[width=0.8\linewidth]{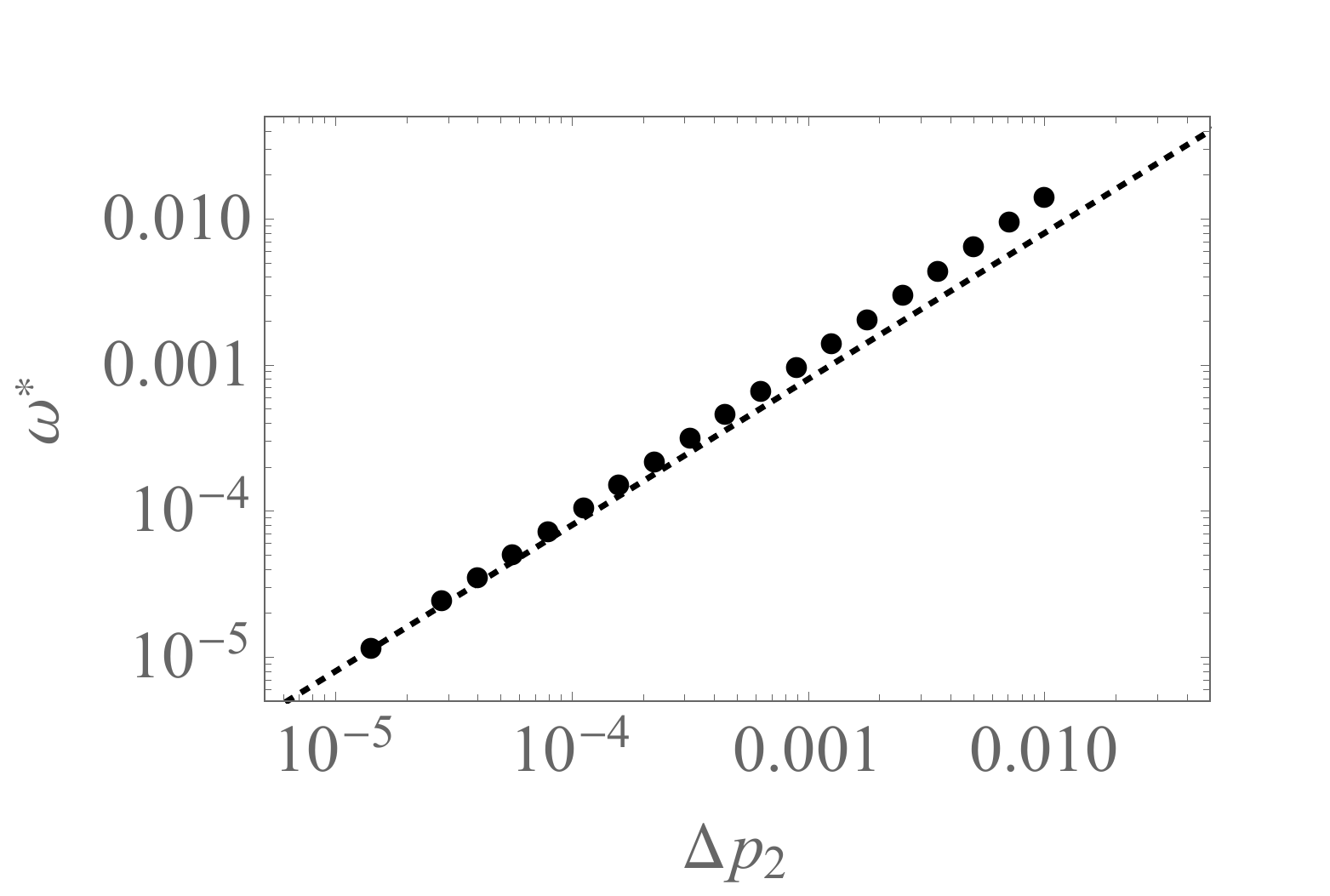}
			\end{minipage}
			\begin{minipage}[t]{0.48\linewidth}
				\centering (c) \par\smallskip
				\includegraphics[width=0.8\linewidth]{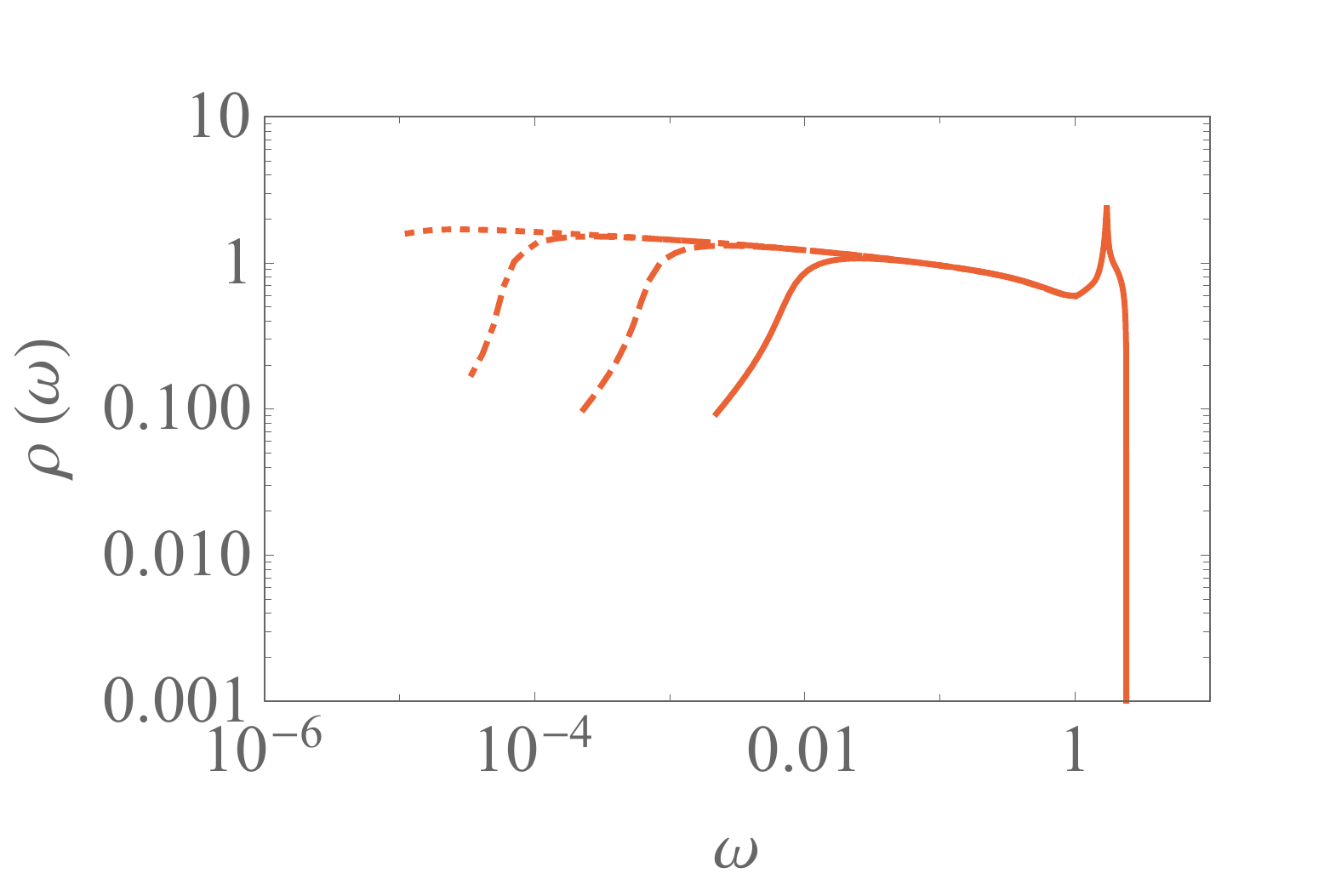}
			\end{minipage}
			\hspace{0.01cm}
			\begin{minipage}[t]{0.48\linewidth}
				\centering (d) \par\smallskip
				\includegraphics[width=0.8\linewidth]{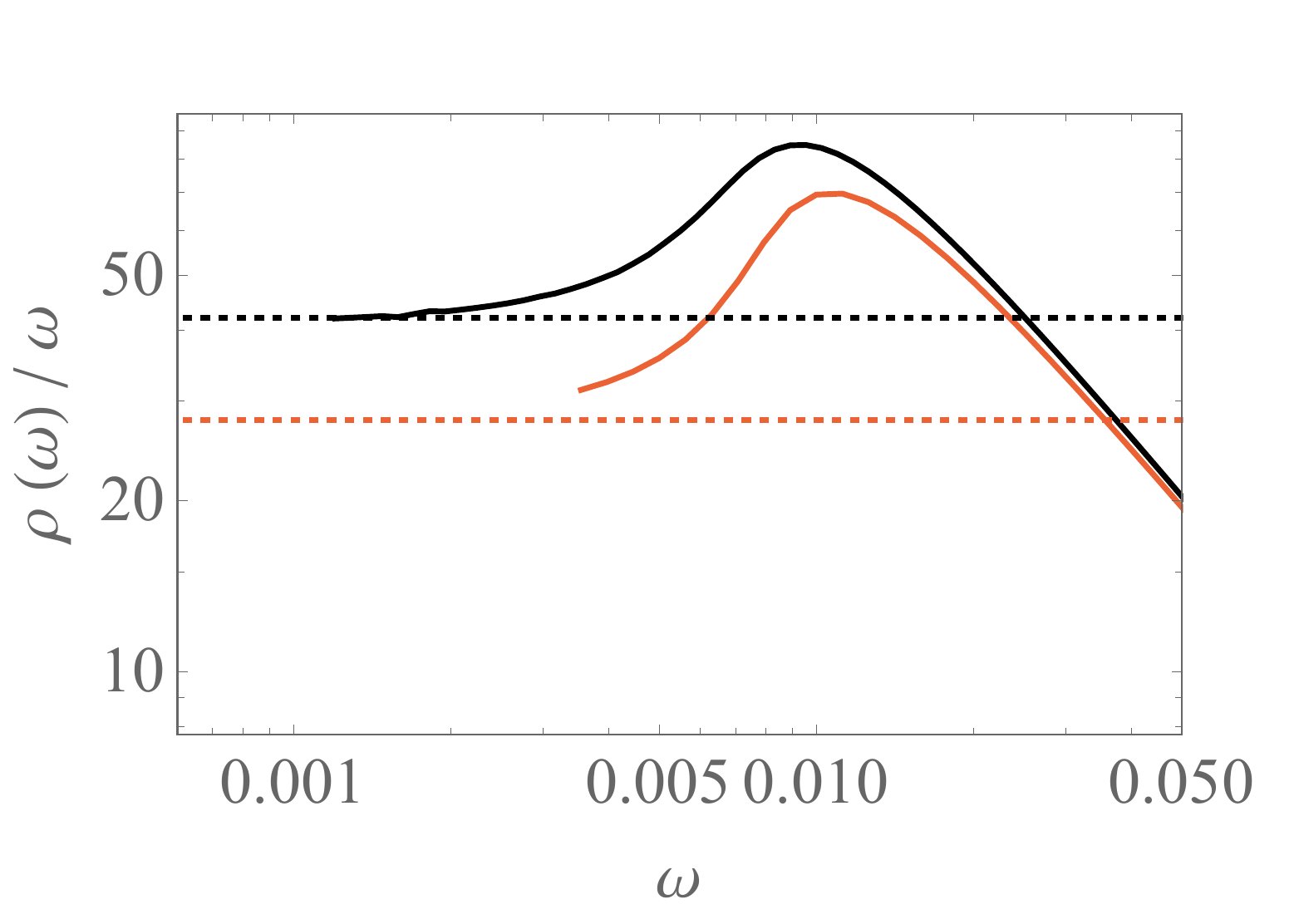}
			\end{minipage}
			\caption{
				Density of states $\rho$ as a function of frequency $\omega$ for $p_a=1$
				((a), near jamming), and $p_a=0.9$ ((c), near rigidity percolation), with
				$\Delta \tilde{p}=10^{-2}$ (solid lines), $10^{-3}$ (dashed), $10^{-4}$
				(dot-dashed) and $10^{-5}$ (dotted).
				(b) Crossover frequency $\omega^*$ as a function of $\Delta \tilde{p}$ near
				the jamming point.
				The black dotted line corresponds to Eq.~\eqref{eq:approx_ws}.
				(d) $\rho(\omega) / \omega$ as a function of $\omega$ for
				$\Delta \tilde{p}=10^{-2}$ near jamming ($p_a=1$, black) and rigidity
				percolation ($p_a=0.6$, red). Solid and dotted lines represent the full
				density of states (Eq.~\eqref{eq:rho_full}) and Debye's approximate
				solutions (Eq.~\eqref{eq:rho_Debye}), respectively.
				\label{fig:dos_low_frequency}}
		\end{figure*}

		Let us consider $p_2 \equiv p_b=p_c$ (as in Section~\ref{subsubsec:nearJ}), and
		assume $\Delta \tilde{p}>0$ and $w(\omega)=\omega^2$.
		The density of states can be calculated as an integral of the imaginary part of the
		retarded Green's function,
		\begin{equation}
			\rho (\omega)
				= \frac{\omega}{\pi \, N_c} \sum_{\bm{q}} \, \text{Tr} \, \text{Im}
					\left[\mathcal{G} ({\bm{q}}, \omega) \right].
			\label{eq:rho_full}
		\end{equation}
		To calculate $\rho$ for given $p_a$, $p_2$ and $\omega$, we first need solve
		the dynamic CPA equations~\eqref{eq:EMT_equations_2} for $k_a$ and $k_2$,
		and use the solution in $\mathcal{G} ({\bm{q}}, \omega)$ in Eq.~\eqref{eq:rho_full}.
		At very small frequency, we expect the density of states display linear behavior in
		two dimensions,
		\begin{equation}
			\rho_D (\omega)
				= \left( \frac{3\sqrt{3}}{4 \pi \, {c_T}^2} + \frac{3\sqrt{3}}{4 \pi \, {c_L}^2}\right)
					\omega,
			\label{eq:rho_Debye}
		\end{equation}
		where ${c_T}^2 = G$ and ${c_L}^2 = B+G$ are the transverse and longitudinal
		sound velocities in 2$D$.
		Equation~\eqref{eq:rho_Debye} can be rewritten as,
		\begin{equation}
			\frac{4 \pi}{3 \sqrt{3}} \frac{\rho_D \, G}{\omega}
				= 1 + \frac{1}{1+B/G}.
		\end{equation}
		Using $B=s_a k_a + 2 s_b k_b$ and $G=2 r_b k_b$, with $s_a=3/4$, $s_b=9/4$,
		and $r_b =9/8$, we find,
		\begin{align}
			\frac{4 \pi}{3 \sqrt{3}} \frac{\rho_D \, G}{\omega}
				& = 1 + \frac{r_b}{r_b + s_b+\displaystyle\frac{s_a}{2} \frac{k_a}{k_b}}
				\nonumber \\ &
				\approx 1 +
					\begin{cases}
						\displaystyle\frac{2\, r_b }{s_a} \frac{k_b}{k_a} \left(\approx 0\right),
							& \text{near } p_a=1 \, (J),
						\\
						\displaystyle\frac{r_b}{r_b + s_b} \left(= \frac{1}{3}\right),
							& \text{near } p_a=0,
					\end{cases}
		\end{align}
		so that the slope of the Debye density of states can vary by a factor of $4/3$
		along the rigidity percolation transition line.
		
		Figure~\ref{fig:dos_low_frequency} shows our EMT solutions for the density of
		states as a function of frequency for $p_a=1$ ((a), near jamming), and $p_a=0.9$
		((c), near rigidity percolation), with $\Delta \tilde{p}=10^{-2}$ (solid line), $10^{-3}$
		(dashed), $10^{-4}$ (dot-dashed) and $10^{-5}$ (dotted).
		As in jammed systems~\cite{SilbertNag2005}, the density of states is nearly
		constant at very small frequencies, down to the crossover frequency
		$\omega^* \sim \Delta \tilde{p}$.
		For $\omega < \omega^*$, the density of states is well-described by Debye
		elasticity, with $\rho \propto \omega$.
		For $p_a=1$ (near jamming) we calculate $\omega^*$ by solving the complex
		CPA equation
		\begin{equation}
			k_2
				= \frac{p_2 - h_2 (\omega^*)}{1-h_2 (\omega^*)},
		\end{equation}
		for real $k_2^*$ and $\omega^*$, using $k_2 = k_2^* (1-i)$.
		In this way, $\omega^*$ is defined as the smallest frequency for which the CPA
		solution satisfies the relation: $\text{Re}(k_2)=-\text{Im}(k_2)$.
		Using Eq.~\eqref{eq:dyn_k2_nearJ}, assuming
		$4(s+\nu c_J) v_2 \omega^2>|\Delta \tilde{p}|^2$ (so that $\text{Im}(k_2) \neq 0$),
		we find
		\begin{align}
			& \text{Re} (k_2)
				\approx \frac{\Delta \tilde{p}}{2\,(s+\nu \, c_J)},
			\nonumber \\ &
			\text{Im} (k_2)
				\approx - \frac{\sqrt{4\,(s+\nu \, c_J \, v_2 \, \omega^2)
					- |\Delta \tilde{p}|^2}}{2(s+\nu \, c_J)},
		\end{align}
		so that,
		\begin{equation}
			\omega^*
				\approx \sqrt{\frac{1}{2 \,(s + \nu \, c_J) v_2}} \, |\Delta \tilde{p}|,
			\label{eq:approx_ws}
		\end{equation}
		which is consistent with our direct numerical calculations.

		Figure~\ref{fig:dos_low_frequency}b shows numerical solutions for $\omega^*$
		as a function of $\Delta \tilde{p} \equiv \Delta p_2$ (black points), indicating linear
		behavior in the vicinity of the jamming point.
		The black dotted line corresponds to Eq.~\eqref{eq:approx_ws}.
		Even though we have not performed an explicit numerical calculation near the
		rigidity percolation line, Fig~\ref{fig:dos_low_frequency}c indicates that $\omega^*$
		also increases linearly with $\Delta \tilde{p}$ in this case.
		Although jamming and rigidity percolation display the same crossover behavior
		to isostaticity, the slope of their low-frequency (Debye) density of states can be
		different for two reasons.
		On the one hand, the longitudinal sound velocity vanishes at $\Delta \tilde{p}=0$
		for rigidity percolation, but not for jamming, which results in a larger contribution
		in the former from the second term of Eq.~\eqref{eq:rho_Debye} for RP.
		On the other hand, the slope of the shear modulus ($G/\Delta \tilde{p}$) decreases
		with $p_a$ (see Fig. 3a of the main text), which results in a smaller transverse
		sound velocity for jamming, and consequently a larger slope of the low-frequency
		density of states.
		
\section{Simulation Setup and elastic moduli of the DLM}
	\label{sec:simulation}
	
	Here we make a few remarks about the simulation setup, and show results
	for the elastic moduli of the DLM in one particular case.
	
	We study the effects of disorder in our model networks by performing numerical
	simulations as well as analytic calculations based on a coherent potential
	approximation.
	In our numerical simulations, we generate our model networks on a computer
	with up to  $100^2$ cell for the honeycomb lattice and up to $10^3$ cells for the
	diamond lattice.
	As mentioned above, all nearest neighboring sites are connected by Hookean
	springs with unit spring constant and probability $p_a$, and next nearest
	neighboring sites are connected with Hookean springs also with unit spring
	constant and probability $p_b$ and $p_c$.
	We implement periodic boundary conditions across all boundaries to moderate
	boundary effects.

	We calculate the elastic response of the these networks by decomposing the
	elastic displacement into an affine and a non-affine part,
	$\bm{u}_i = \eta \, \bm{r}_i + \delta \bm{u}_i$, and then relaxing the non-affine part
	$\delta \bm{u}_i = \bm{u}_i - \eta \, \bm{r}_i$ with a standard conjugate gradient
	algorithm.
	To compute the shear modulus, for example, we set all the diagonal elements of
	$\eta$ equal to zero and all off-diagonal elements to $\gamma$, where $\gamma$
	is the magnitude of the deformation.
	We use $\gamma = 10^{-2}$ throughout.
	In addition to the elastic moduli, we compute the phonon density of states by
	diagonalizing the dynamical matrix of our networks numerically.
	All quantities are computed for several network realizations for given $p_a$, $p_b$
	and $p_c$ and then arithmetically averaged over these realizations.

	\begin{figure}[!ht]
		\centering
		\includegraphics[width=0.8\linewidth]{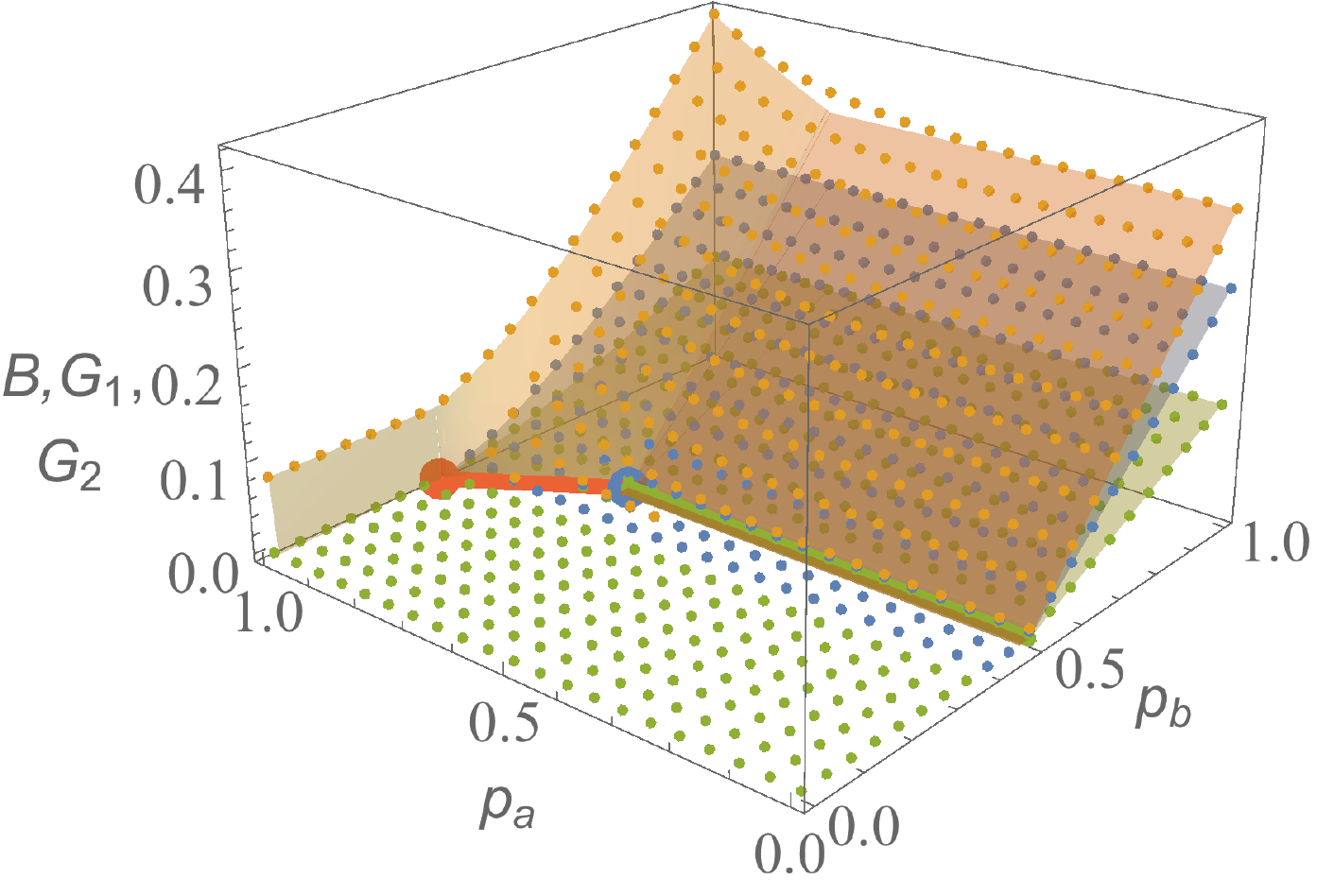}
		\caption{
			Simulation (points) and EMT solutions (surfaces) for the bulk modulus
			$B$ (yellow), and the two shear moduli of the diamond lattice, $G_1$
			(green) and $G_2$ (blue), as a function of $p_a$ and $p_b$
			for $p_c=0$.
			Note that this result is qualitatively similar to that for the 2D HLM (see
			Fig.~\ref{fig:plots} in Section~\ref{sec:cpa} and Fig. 2 of the main text).
			\label{fig:moduli_diamond}}
	\end{figure}
	
	Figure~\ref{fig:moduli_diamond} shows simulations (points) and EMT solutions
	(surfaces) for the bulk modulus $B$ (yellow) and the two shear moduli
	of the diamond lattice, $G_1$ (green) and $G_2$ (blue), as a
	function of $p_a$ and $p_b$ with $p_c=0$.
	This sample example indicates that the main results for the 3D DLM are qualitatively
	similar to those for the 2D HLM.
	
\section{An Isotropic Model with $p_b^J =0$}

\begin{figure}[!ht]
	\centering
	\includegraphics[width=0.4\linewidth]{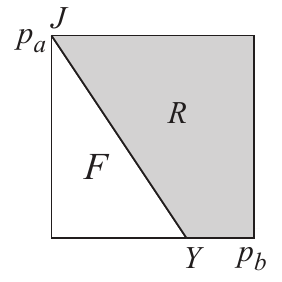}
	\caption{Phase diagram for a model with the jamming point $J$ at $p_a=1$, $p_b = 0$.  Unlike the models based on the honeycomb and diamond lattices, there is no line of transitions from the floppy phase to one with $B>0$ and $G=0$. The rigidity  percolation line $JY$ connects the point $J$ to the point $Y=(p_a=0, p_b = D/\tilde{z}_b)$.}	
	\label{fig:new-phase-dia}
\end{figure}

	Our model uses the facts that the under-coordinated $NN$ honeycomb and diamond lattices have nonzero bulk moduli and they develop a nonvanishing shear modulus at a jamming critical point $J=(1,p_b^J)$ at a critical concentration $p_b^J$ of $NNN$ springs.  Though this model captures the essential features of jamming, it does have the undesirable property that the bulk modulus is nonzero for all $p_a = 1$, even when $p_b <p_b^J$, and, as a result, there is a first-order transition from the floppy phase with $B=G=0$ to the critical line $p_a = 1$, $0<p_b < p_b^J$. This deficiency is eliminated in a generalization of the of the isotropic EMT considered in Refs.~\cite{WyartMah2008,Wyart2010,DuringWya2013}. We start with a random $a$-lattice at criticality with $\tilde{z}_a = D = z_a/2$ and then add extra ``$NNN$" bonds creating a $b$-lattice.  As before, we occupy the bonds in the two lattices with springs respective probabilities $p_a$ and $p_b$.  The EMT equations for this model are identical to those of the honeycomb and diamond based lattices [Eqs.~(2) to (4) in the main article] with $m = 1$ because we do not divide the lattice into multi-site cells and $\tilde{z}_a=D$:
\begin{align}
	D h_a + \tilde{z}_b h_b & = D \\
	h_\alpha & = \frac{1}{\tilde{z}_\alpha N} \sum_\qv \text{Tr} k_\alpha K_\alpha D .
	\label{Eq:h-alpha1}
\end{align}
The Maxwell criterion for the jamming tansition is then
\begin{equation}
	D p_a + \tilde{z}_b p_b = D.
\end{equation}
At the jamming point $p_a = p_a^J = 1$, and thus, $p_b^J = 0$, i.e., as shown in Fig.~\ref{fig:new-phase-dia}, $J$ is now located at the upper left-hand corner of the $p_a-p_b$ plane, and there is no phase transition from the floppy phase to one with $B>0$ and $G=0$ except at the point $J$.  The $RP$ line for $p_a <1$ is then
\begin{equation}
	\Delta \tp = p_b - \nu \Delta p_a ,
\end{equation}
where $\nu = D/\tilde{z}_b$ and $\Delta p_a = 1-p_a$, implying that the $RP$ transition at $p_a = 0$ is at $p_b = p_b^Y = \nu$. For small $k_b/k_a$, the equations for $h_a$ and $h_b$ are identical to Eqs.~(\ref{eq:ka_cpa_nearJ}) and (\ref{eq:k2_cpa_nearJ}) with $p_2^J = 0$.  The equations for $k_a$ and $k_b$ are the same as Eqs.~(\ref{eq:k2_cpa_nearJ}) to (\ref{eq:ka_nearJ_loww}) with $\Delta p_a = 1-p_a$ and $p_2 = p_b$.

\clearpage

\end{document}